\documentclass[journal,a4paper]{IEEEtran}
\IEEEoverridecommandlockouts

\usepackage[utf8]{inputenc} 
\usepackage[T1]{fontenc}
\usepackage{url}
\usepackage{ifthen}
\usepackage{cite}
\usepackage[cmex10]{amsmath} 
\usepackage{amssymb}
\usepackage{graphicx}
\usepackage{epstopdf}
\usepackage{epsfig}
\usepackage{multirow}
\usepackage{amsthm}
\usepackage{comment}
\usepackage{caption}
\usepackage{enumitem}
\usepackage{amsmath}
\usepackage{color}
\usepackage{bbm}
\usepackage{algorithm}
\usepackage{algpseudocode}
\usepackage{csquotes}
\usepackage{subfiles}
\usepackage{caption, multirow, makecell}
\usepackage{array}
\usepackage{caption}
\usepackage{afterpage}
\usepackage{dblfloatfix}
\usepackage{hyperref}
\usepackage{graphicx}
\usepackage{balance}
\usepackage{subcaption}
\newtheorem{thm}{Theorem}

\newtheorem{example}{Example}

\newtheorem{defn}{Definition}

\newtheorem{rem}{Remark}

\def\BibTeX{{\rm B\kern-.05em{\sc i\kern-.025em b}\kern-.08em
		T\kern-.1667em\lower.7ex\hbox{E}\kern-.125emX}}

\begin{document}

\title{Optimal Placement Delivery Arrays from $t$-Designs with Application to Hierarchical Coded Caching \\}

\author{\IEEEauthorblockN{Rashid Ummer N.T. and B. Sundar Rajan} \\
	\IEEEauthorblockA{\textit{Department of Electrical Communication Engineering} \\
		\textit{Indian Institute of Science}\\
		Bangalore, India \\
		\{rashidummer,bsrajan\}@iisc.ac.in}
}

\maketitle

\begin{abstract}
Coded caching scheme originally proposed by Maddah-Ali and Niesen (MN) achieves an optimal transmission rate $R$ under uncoded placement but requires a subpacketization level $F$ which increases exponentially with the number of users $K$ where the number of files $N \geq K$. Placement delivery array (PDA) was proposed as a tool to design coded caching schemes with reduced subpacketization level by Yan \textit{et al.} in \cite{YCT}. This paper proposes two novel classes of PDA constructions from combinatorial $t$-designs that achieve an improved transmission rate for a given low subpacketization level, cache size and number of users compared to existing coded caching schemes from $t$-designs. A $(K, F, Z, S)$ PDA composed of a specific symbol $\star$ and $S$ non-negative integers corresponds to a coded caching scheme with subpacketization level $F$, $K$ users each caching $Z$ packets and the demands of all the users are met with a rate $R=\frac{S}{F}$. For a given $K$, $F$ and $Z$, a lower bound on $S$ such that a $(K, F, Z, S)$ PDA exists is given by Cheng \textit{et al.}  in \cite{MJXQ} and by Wei in \cite{Wei} . Our first class of proposed PDA achieves the lower bound on $S$. The second class of PDA also achieves the lower bound in some cases. From these two classes of PDAs, we then construct hierarchical placement delivery arrays (HPDA), proposed by Kong \textit{et al.} in \cite{KYWM}, which characterizes a hierarchical two-layer coded caching system. These constructions give low subpacketization level schemes.

\end{abstract}
\begin{IEEEkeywords}
 Coded caching, combinatorial design, subpacketization, placement delivery array, hierarchical coded caching.
\end{IEEEkeywords}

\section{Introduction}

Wireless data traffic has increased phenomenally in the past years, more than doubling on average every second year and  is forecast to grow 20–30 percent per year in the next few years \cite{Eri}. 
In cache aided communication, by taking advantage of the memories distributed across the networks, some of the peak time traffic has shifted to off-peak times. The novel centralized coded caching scheme proposed in \cite{MaN} (referred to as MN scheme), achieves a \textit{global caching gain} in addition to the \textit{local caching gain} achieved in uncoded caching schemes. The central idea is to jointly optimize the content \textit{placement phase} during off-peak times and the \textit{delivery phase} during peak time, ensuring that a single coded multicast transmission serves multiple users. 
The network model consists of a single server having access to a library of $N$ equal-length files connected through a shared error-free link to $K$ users each possessing a cache of size equal to $M$ files. For this network model, the transmission rate $R$  achieved by the MN scheme is shown to be optimal under uncoded data placement where $N \geq K$ \cite{YMA}. To obtain the optimal rate in the MN scheme, each file has to be split into  $F$ number of packets, referred to as the subpacketization level,  which grows exponentially with the number of users. Thus for practical implementations, it is desirable to have coded caching schemes with reduced subpacketization levels.

A combinatorial structure called Placement Delivery Array (PDA) which describes both placement and delivery phases in a single array was proposed by Yan \textit{et al.} in \cite{YCT}. PDAs can be designed for a reduced subpacketization level than that of the MN scheme at the expense of an increase in the transmission rate. Various coded caching schemes with reduced subpacketization level such as schemes using the linear block codes \cite{TaR}, the strong edge coloring of bipartite graphs \cite{Yan}, the Ruzsa-Szeméredi graphs \cite{Ruz}, the projective space \cite{Pra}, the combinatorial designs \cite{SSP}, \cite{SSParx}, \cite{Li} and \cite{MACC_arx}, the hypergraph theoretical approach \cite{SCYG}  and so on, have been proposed. Shanmugam \textit{et al.} in \cite{Ruz2} has pointed out that the constructions in \cite{TaR}, \cite{Yan}, \cite{Ruz}, \cite{Pra} and\cite{SCYG} can be represented by PDAs.

Characterizing the relationship between  $R$ and  $F$ is a challenging problem. None of the above PDA constructions have discussed the optimality of the coded caching scheme obtained. Jin \textit{et al.} in \cite{Jin} formulated an optimization problem to minimize rate under subpacketization constraint, cache memory constraint and file partition constraint which is an NP-Hard problem. For a coded caching problem with a given $K$, $F$ and $Z$ which can be represented as a PDA, a lower bound on $S$ (equivalently on $R$) is given by Cheng \textit{et al.} in \cite{MJXQ} and by Niu \textit{et al.} in \cite{Niu}. Optimization of PDAs is also discussed by Ruizhong Wei in \cite{Wei} by giving another lower bound on $S$. 

A hierarchical two-layer coded caching scheme consisting of a single server connected to multiple mirror sites and each mirror site connected with a distinct set of users, in which both mirror sites and users having cache memories was introduced in \cite{KNMD}. Coded caching schemes for such a two-layer network under different settings are also discussed in \cite{KYWM}, \cite{ZZWXL} and \cite{WWCY}. A combinatorial structure called hierarchical placement delivery array (HPDA) which describes both placement and delivery phases for a hierarchical coded caching system was proposed by Yun \textit{et al.} in \cite{KYWM}.

\noindent \textit{Notations}: For any set $\mathcal{A}$, $|\mathcal{A}|$ denotes the cardinality of $\mathcal{A}$. 
For a set $\mathcal{A}$ and a positive integer $i \leq |\mathcal{A}|$,  $\binom{\mathcal{A}}{i}$ denotes all the $i$-sized subsets of $\mathcal{A}$. For sets $\mathcal{A} \text{ and }\mathcal{B}$, $\mathcal{A}\textbackslash \mathcal{B}$ denotes the elements in $\mathcal{A}$ but not in $\mathcal{B}$ and $\emptyset$ denotes the empty set.
\subsection{Contributions} 
The contributions of this paper are summarized below:
\begin{itemize}
	\item  We present two novel classes of PDA constructions from combinatorial $t$-designs which give improved transmission rates compared to the existing schemes from  combinatorial $t$-designs for a given subpacketization level, cache size and number of users.
	\item The first class of PDA achieves optimal rate with respect to the lower bound on $S$ for a given low subpacketization level, cache size and number of users. 
	\item The second class of PDA is also shown to be optimal in some cases.
	\item From these new classes of PDAs, we then construct hierarchical placement delivery arrays (HPDA), characterizing a hierarchical two-layer caching system that has a low subpacketization level compared to the existing schemes. 
	\end{itemize}

\subsection{Organization}
The rest of the paper is organized as follows. In Section II, we briefly review the preliminaries of PDA, combinatorial t-designs and some known results and bounds. The two classes of novel PDA constructions from $t$-designs are discussed in Section III.  In Section IV, the performance analysis of the proposed constructions is carried out by comparing them with existing schemes from t-designs and also by discussing the optimality of proposed PDAs. In Section V, HPDA construction from the proposed PDAs and its performance analysis are discussed. Section VI concludes the paper.
 
\section{Preliminaries}
\subsection{Placement Delivery Array (PDA)}
\begin{defn}
	(\hspace{1sp}\cite{YCT}) For positive integers $K, F, Z$ and $S$, an $F \times K$ array $\mathbf{P}=(p_{j,k})$, $j \in [0,F)$ and $k \in [0,K)$, composed of a specific symbol $\star$ and $S$ non-negative integers $0,1,\ldots, S-1$, is called a $(K, F, Z, S)$ placement delivery array (PDA) if it satisfies the following three conditions: \\
	\textit{C1}. The symbol $\star$ appears $Z$ times in each column.\\
	\textit{C2}. Each integer occurs at least once in the array.\\
	\textit{C3}. For any two distinct entries $p_{j_1,k_1}$ and $p_{j_2,k_2}$,\\ $p_{j_1,k_1}=p_{j_2,k_2}=s$ is an integer only if (\textit{a}) $j_1 \neq j_2$, $k_1 \neq k_2$, i.e., they lie in distinct rows and distinct columns, and (\textit{b}) $p_{j_1,k_2}=p_{j_2,k_1}=\star$.
\end{defn} 
\begin{defn}
	(\hspace{1sp}\cite{YCT}) An array $\mathbf{P}$ is said to be a $g$-regular $(K, F, Z, S)$ PDA, $g-(K, F, Z, S)$ PDA or $g-$ PDA for short, if it satisfies \textit{C1}, \textit{C3}, and the following condition \\
	\textit{C2'}. Each integer appears $g$ times in $\mathbf{P}$ where $g$ is a constant.
\end{defn}
\begin{thm}
	(\hspace{1sp}\cite{YCT}) For a given $(K, F, Z, S)$ PDA $\mathbf{P}=(p_{j,k})_{F \times K}$, a $(K,M,N)$ coded caching scheme can be obtained with subpacketization $F$ and $\frac{M}{N}=\frac{Z}{F}$ using Algorithm \ref{alg1}. For any demand $\mathbf{d}$, the demands of all the users are met with a transmission load of $R=\frac{S}{F}$.
\end{thm}
\begin{algorithm}
	\renewcommand{\thealgorithm}{1}
	\caption{Coded caching scheme based on PDA \cite{YCT}}
	\label{alg1}
	\begin{algorithmic}[1]
		\Procedure{Placement}{$\mathbf{P},\mathcal{W}$}       
		\State Split each file $W^n$ in $\mathcal{W}$ into $F$ packets: $W^n =\{W^n_j: j \in [0,F)\}$
		\For{\texttt{$k \in 0,1,\ldots,K-1$}}
		\State  $\mathcal{Z}_k$ $\leftarrow$ $\{W^n_{j}: p_{j,k}=\star, \forall n \in [N]\}$
		\EndFor
		\EndProcedure
		
		\Procedure{Delivery}{$\mathbf{P},\mathcal{W},\mathbf{d}$} 
		\For{\texttt{$s = 0,1,\ldots, S-1$}}
		\State Server sends $\underset{p_{j,k}=s, j\in [0,F),k\in[0,K)}{\bigoplus}W^{d_k}_j$
		\EndFor    
		\EndProcedure
	\end{algorithmic}
\end{algorithm}

In a $(K, F, Z, S)$ PDA $\mathbf{P}$, the column indices represent users and row indices represent packets. For any $k \in [0,K)$ , the $j^{th}$ packet of all the files is placed in  $k^{th}$ user's cache only if $p_{j,k}=\star$. Condition $C1$ guarantees that each user has the same memory size and memory ratio $\frac{M}{N}= \frac{Z}{F}$. In the delivery phase, corresponding to each integer $s$ in the PDA, the server sends a linear combination of the requested packets indicated by that integer. Condition $C3$ ensures the decodability of packets of requested file by each user. By condition $C2$, the number of transmissions by the server is exactly $S$ and therefore, the transmitted rate $R=\frac{S}{F}$.
\begin{rem}
	If the number of $\star$ is same in every row (say $Z'$) in a given $(K,F,Z,S)$ PDA $\mathbf{P}$, then $Z'=\frac{Total\hspace{0.1cm} number\hspace{0.1cm} of\hspace{0.1cm} \star s\hspace{0.1cm} in\hspace{0.1cm} \mathbf{P}}{number\hspace{0.1cm} of \hspace{0.1cm}rows}=\frac{KZ}{F}$. Therefore, the transpose of $\mathbf{P}$ gives another $(F, K, Z', S)$ PDA $\mathbf{P'}$ which corresponds to a coded caching scheme with same memory ratio $\frac{M}{N}$ and rate $R=\frac{S}{K}$.
\end{rem}
\begin{table*}[!b]
	\centering
	\caption{Summary of known PDA constructions from $t$-designs}
	\begin{tabular}{| c | c | c | c | c| c|}
		\hline
		\rule{0pt}{4ex}
		\makecell{Schemes and parameters} & \makecell{Number of users\\ $K$} & \makecell{Caching ratio \\ $\frac{M}{N}=\frac{Z}{F}$} & \makecell{Subpacketization \\ $F$} & \makecell{Number of integers\\ $S$}& \makecell{Rate\\ $R=\frac{S}{F}$}  \\ [4pt]		
		\hline
		\rule{0pt}{3.5ex}
		\makecell{Scheme I in \cite{SSP} \\  $t-(v, k, 1)$ design} & $\binom{v}{t-1}$ & $1-\frac{\binom{k}{t}(v-t+1)}{\binom{v}{t}k}$ & $\frac{\binom{v}{t}k}{\binom{k}{t}}$ & $\binom{k-1}{t-1}v$ & $\frac{\binom{k-1}{t-1}\binom{k}{t}v}{\binom{v}{t}k}$ \\	
		\hline
		\rule{0pt}{3.5ex}
		\makecell{Scheme II in \cite{SSParx} \\ $t-(v, k, 1)$ design} & $v$ & $1-\frac{t}{v}$ & $\binom{v}{t}$ & $\binom{v}{t-1}$ & $\frac{t}{v-t+1}$ \\		
		\hline
		\rule{0pt}{3.5ex}
		\makecell{Theorem 5 in \cite{Li} \\  $t-(v, k, 1)$ design \\ for $i \in [1,t-1]$ }  & $\frac{\binom{v}{t}\binom{k}{i}}{\binom{k}{t}}$ & $1-\frac{\binom{k-i}{t-i}}{\binom{v}{t-i}}$ & $\binom{v}{t-i}$ & $\binom{k-i}{t-i}\binom{v}{i}$ & $\frac{\binom{v}{i}\binom{k-i}{t-i}}{\binom{v}{t-i}}$ \\
		\hline
		\rule{0pt}{3.5ex}
		\makecell{Theorem 6 in \cite{Li} \\  simple $t-(v, k, \lambda)$ design \\ with $k \leq 2t$ }  & $\binom{v}{t}$ & $1-\frac{\binom{k}{t}}{\binom{v}{t}}$ & $\frac{\lambda\binom{v}{t}}{\binom{k}{t}}$ & $\binom{v}{k-t}$ & $\frac{\binom{v}{k-t}\binom{k}{t}}{\lambda \binom{v}{t}}$ \\
		\hline
		\rule{0pt}{3.5ex}
		\makecell{Theorem 1 in \cite{MACC_arx} \\  $t-(v, k, 1)$ design \\ for $i \in [0,v-k]$ }  & $\frac{\binom{v}{t}}{\binom{k}{t}}$ & $1-\frac{\binom{v-k}{i}}{\binom{v}{i}}$ & $\binom{v}{i}{\binom{k}{t}}$ & $\binom{v}{t+i}-\frac{\binom{v}{t}\binom{k}{t+i}}{\binom{k}{t}}$ & $\frac{\binom{v}{t+i}}{\binom{v}{i}\binom{k}{t}}-\frac{\binom{v}{t}\binom{k}{t+i}}{{\binom{k}{t}}^2\binom{v}{i}}$ \\
		\hline
	\end{tabular}
	
	\label{tab:pda}
\end{table*} 
\subsection{Optimal PDA}
\begin{defn}[Optimal PDA]
	For a coded caching problem with given number of users $K$, number of files $N$, subpacketization level $F$ and cache memory $M$ which can be represented as a $(K,F,Z,S)$ PDA $\mathbf{P}$, the PDA is said to be an \textit{optimal PDA} if the number of distinct integers $S$ in $\mathbf{P}$ is the least possible integer such that a $(K,F,Z,S)$ PDA $\mathbf{P}$ exists. 
\end{defn}
Formulating optimality of a PDA in above fashion is also done in \cite{Wei}. For a given number of users $K$, subpacketization level $F$ and each user caching $Z$ packets of every file which can be represented as a PDA, a lower bound for $S$ is given in \cite{MJXQ}, \cite{Niu} and \cite{Wei}.
\begin{thm}\label{thm:MJXQ}
	(\hspace{1sp}\cite{MJXQ},\cite{Niu}) Given any positive integers $K,F,Z$ with $0<Z<F$, if there exists a $(K,F,Z,S)$ PDA, then 
	\begin{equation}\label{eq:MJXQ}
		\begin{split}
			& S \geq \left\lceil \frac{(F-Z)K}{F}\right\rceil + \left\lceil \frac{F-Z-1}{F-1} \left\lceil \frac{(F-Z)K}{F}\right\rceil \right\rceil \\ & + ...+ \left\lceil \frac{1}{Z+1} \left\lceil \frac{2}{Z+2} \left\lceil ...\left\lceil \frac{(F-Z)K}{F} \right\rceil ... \right\rceil \right\rceil \right\rceil .
		\end{split}
	\end{equation}	
\end{thm} 
\begin{thm}\label{thm:Wei}
	(\hspace{1sp}\cite{Wei}) In any $(K,F,Z,S)$ PDA,
	\begin{equation}\label{eq:Wei}
		S \geq \left\lceil \frac{K(F-Z)}{Z+1}\right\rceil .
	\end{equation}	
\end{thm}
One necessary condition to hold equality in (\ref{eq:Wei}) is 
\begin{equation*}
	K > \frac{(Z+1)(F-Z-1)}{F-Z}.
\end{equation*}
A PDA which satisfies the lower bound in Theorem~\ref{thm:Wei} with equality is defined as \textit{Restricted PDA (RPDA)} in \cite{Wei}. We call a PDA as \textit{minimal load PDA}, if it satisfies the lower bound in Theorem~\ref{thm:MJXQ} with equality. An RPDA is an optimal PDA, but an optimal PDA may not be an RPDA. Similarly, a \textit{minimal load PDA} is an optimal PDA, but an optimal PDA may not be a \textit{minimal load PDA}. In general, the lower bound for $S$ in (\ref{eq:MJXQ}) is a tighter bound than in (\ref{eq:Wei}). (Proof given in \textit{Appendix A}). That is, an RPDA is a \textit{minimal load PDA} but a \textit{minimal load PDA} may not be an RPDA. Examples \ref{eg:opt1}, \ref{eg:opt2} and \ref{eg:opt3} illustrate this. 
\begin{example}\label{eg:opt1}
	For $K=6,F=8,Z=5$, by  Theorem~\ref{thm:MJXQ}, $S \geq 3+1+1=5$. The following $(6,8,5,5)$ PDA is therefore an \textit{optimal PDA}.  In this case, (\ref{eq:Wei}) gives $S \geq 3$.  
	\begin{equation*}
		\mathbf{P} = \begin{bmatrix}
			1 & \star & \star & \star & 4 & \star \\
			2 & 4 & \star & \star & \star & 5 \\
			\star & 1 & 2 & \star & \star & \star \\
			3 & \star & 4 & \star & \star & \star \\
			\star & 3 & \star & 2 & \star & \star \\
			\star & \star & 5 & 1 & 3 & \star \\
			\star & \star & \star & \star & 2 & 1 \\
			\star & \star & \star & 4 & \star & 3 \\
		\end{bmatrix}
	\end{equation*}
\end{example}
\begin{example}\label{eg:opt2}
	For $K=6,F=4,Z=2$, by  Theorem~\ref{thm:MJXQ}, $S \geq 3+1=4$. The following $(6,4,2,4)$ PDA is therefore an \textit{optimal PDA}. In this case, it is same as the lower bound given by (\ref{eq:Wei}), $S \geq 4 $ . 
	\begin{equation*}
		\mathbf{P} = \begin{bmatrix}
			\star & \star & \star & 1 & 2 & 3 \\
			\star & 1 & 2 & \star & \star & 4 \\
			1 & \star & 3 & \star & 4 & \star \\
			2 & 3 & \star & 4 & \star & \star \\
		\end{bmatrix}
	\end{equation*}
\end{example} 
\begin{example}\label{eg:opt3}
The following $(7,7,4,6)$ PDA is an \textit{optimal PDA} \cite{Wei}. For this PDA, by  Theorem~\ref{thm:MJXQ}, $S \geq 3+1+1=5$. In this case, (\ref{eq:Wei}) gives $S \geq 5$.  
\begin{equation*}
	\mathbf{P} = \begin{bmatrix}
		1 & \star & \star & \star & 2 & \star & 4 \\
		\star & 1 & \star & \star & \star & 2 & 6 \\
		\star & \star & 1 & \star & \star & 5 & 3 \\
		\star & \star &\star & 1 & 6 & 4 & \star \\
		5 & 3 & 2 & \star & \star & \star & \star \\
		6 & 4 & \star & 2 & \star & \star & \star \\
		\star & \star & 4 & 3 & 5 & \star & \star \\
	\end{bmatrix}
\end{equation*}
\end{example}

\subsection{Combinatorial t-design}
Some basic useful definitions and properties  from combinatorial design are described in this subsection. For further reading on this, readers can refer to \cite{Stin}.

\begin{defn}[Design $(\mathcal{X}, \mathcal{A})$]
	A design is a pair  $(\mathcal{X}, \mathcal{A})$ such that the following properties are satisfied: \\
	\textit{1}. $\mathcal{X}$ is a set of elements called points, and\\
	\textit{2}. $\mathcal{A}$ is a collection (i.e., multiset) of nonempty subsets of $\mathcal{X}$ called blocks.
\end{defn}
A design is called \textit{simple design}  if it does not contain repeated blocks.
\begin{defn}[$t-(v, k, \lambda)$ design]
	Let $v,k,\lambda$ and $t$ be positive integers such that $v>k\geq t$. A $t-(v, k, \lambda)$ design is a design $(\mathcal{X}, \mathcal{A})$ such that the following properties are satisfied: \\
	\textit{1}. $|\mathcal{X}|=v$, \\
	\textit{2}. each block contains exactly $k$ points, and \\
	\textit{3}. every set of $t$ distinct points is contained in exactly $\lambda$ blocks.
\end{defn}

The general term $t$-design is used to indicate any $t-(v, k, \lambda)$ design. The construction of $t$-designs is discussed in \cite{Stin}. For notation simplicity, we write blocks and its subsets in the form $abcd$ instead of $\{ a,b,c,d\}$.
 \begin{example}\label{ex:example1}
 	$\mathcal{X}=\{1,2,3,4,5,6,7\}$, \\ $ \mathcal{A}=\{127,145,136,467,256,357,234\}$ is a $2-(7,3,1)$ design. 
 \end{example}
 \begin{example}\label{ex:example2}
	$\mathcal{X}=\{1,2,3,4,5,6\}$, \\ $ \mathcal{A}=\{124,126,134,135,156,235,236,245,346,456\}$ is a $2-(6,3,2)$ design. 
\end{example}
\begin{thm}\label{thm:Stin}
	(\hspace{1sp}\cite{Stin}) Suppose that $(\mathcal{X}, \mathcal{A})$ is a $t-(v, k, \lambda)$ design. Suppose that $\mathcal{Y} \subseteq \mathcal{X}$, where $|\mathcal{Y}|=s\leq t.$ Then there are exactly $\lambda_{s}=\frac{\lambda \binom{v-s}{t-s}}{\binom{k-s}{t-s}}$ blocks in $\mathcal{A}$ that contains all the points in $\mathcal{Y}$. Thus the number of blocks in $\mathcal{A}$, $b=\lambda_{0}=\frac{\lambda \binom{v}{t}}{\binom{k}{t}}$.
\end{thm}

\subsection{Known  $t$-design based coded caching schemes }
Agrawal \textit{et al.} in \cite{SSP} proposed a binary matrix model to construct a coded caching scheme from  $t-(v, k, 1)$ design (Scheme-I) and gave another construction from $t-(v, k, 1)$ design in the longer version \cite{SSParx}. In fact, the binary matrix model can be represented as PDAs.  Li \textit{et al.} in \cite{Li} gave two classes of PDA constructions, one from  $t-(v, k, 1)$ design and another one from $t-(v, k, \lambda)$ design with $k \leq 2t$. In \cite{MACC_arx}, Cheng \textit{et al.} gave a PDA construction from $t-(v, k, 1)$ design which is used as a user delivery array for the multiaccess coded caching scheme they consider. Table \ref{tab:pda} summarizes these schemes. 
\subsection{Hierarchical Placement Delivery Array (HPDA)}
\begin{defn}
	(\hspace{1sp}\cite{KYWM}) For any given positive integers $K_1, K_2, F, Z_1, Z_2$ with $Z_1 < F$, $Z_2 < F $ and any integer sets $S_m$ and $S_{k_1}, k_1 \in [K_1]$, an  $ F \times (K_1+K_1K_2)$ array $\mathbf{Q}= \left( \mathbf{Q^{(0)}},\mathbf{Q^{(1)}},....,\mathbf{Q^{(K_1)}} \right)$, where $\mathbf{Q^{(0)}}=(q_{j,k_1}^{(0)})_{j \in [F], k_1 \in [K_1]}$, is an $F \times K_1$ array consisting of $\star$ and \textit{null}  and $\mathbf{Q^{(k_1)}}=(q_{j,k_2}^{(k_1)})_{j \in [F], k_2 \in [K_2]}$ is an $F \times K_2$  array over  ${\star} \cup S_{k_1}, k_1 \in [K_1]$, is a $(K_1, K_2;F;Z_1,Z_2;S_m,S_1,....,S_{K_1})$ hierarchical placement delivery array  if it satisfies the following conditions: \\
	\textit{B1}. Each column of $\mathbf{Q^{(0)}}$ has $Z_1$ stars.\\
	\textit{B2}. $\mathbf{Q^{(k_1)}}$ is a $(K_2,F,Z_2,|S_{k_1}|)$ PDA for each $ k_1 \in [K_1]$ .\\
	\textit{B3}. Each integer $s \in S_m$ occurs in exactly one subarray $\mathbf{Q^{(k_1)}}$ where $ k_1 \in [K_1]$. And for each $(q_{j,k_2}^{(k_1)})=s \in S_m,j \in [F],k_1 \in [K_1], k_2 \in [K_2]$, $(q_{j,k_1}^{(0)})=\star$. \\
	\textit{B4}. For any two entries $q_{j,k_2}^{(k_1)}$ and $q_{j',k'_2}^{(k'_1)}$, where $k_1 \neq k'_1 \in [K_1]$, $j,j' \in [F]$ and $k_2,k'_2 \in [K_2]$, if $q_{j,k_2}^{(k_1)}=q_{j',k'_2}^{(k'_1)}$ is an integer then
	\begin{enumerate}[label=-]
		
		\item $q_{j',k_2}^{(k_1)}$ is an integer only if $q_{j',k_1}^{(0)}=\star$,
		\item $q_{j,k'_2}^{(k'_1)}$ is an integer only if $q_{j,k'_1}^{(0)}=\star$.\\
	\end{enumerate}
\end{defn}

In a given HPDA $\mathbf{Q}= \left( \mathbf{Q^{(0)}},\mathbf{Q^{(1)}},....,\mathbf{Q^{(K_1)}} \right)$, $\mathbf{Q^{(0)}}$ indicates the data placement in the mirror sites and $\mathbf{Q^{(k_1)}, k_1 \in [K_1]}$ indicates the placement at the users attached to $k_1^{th}$ mirror site (i.e., users in $\mathcal{U}_{k_1}$) and also the delivery from the server and the mirrors to the users attached to $k_1^{th}$ mirror. The corresponding coded caching scheme is explained in Algorithm \ref{alg2}.
\begin{thm}\label{thm:hpda_thm}
	(\hspace{1sp}\cite{KYWM}) Given a $(K_1,K_2;F;Z_1,Z_2;S_m,S_1,..,\\S_{K_1})$  HPDA $\mathbf{Q}= \left( \mathbf{Q^{(0)}},\mathbf{Q^{(1)}},....,\mathbf{Q^{(K_1)}} \right)$, we can obtain an $F$-division $(K_1,K_2;M_1,M_2;N)$ coded caching scheme using Algorithm \ref{alg2} with $\frac{M_1}{N}=\frac{Z_1}{F}$, $\frac{M_2}{N}=\frac{Z_2}{F}$ and transmission load of \\ $R_1=\frac{\left|\underset{k_1=1}{\bigcup^{K_1} S_{k_1}}\right|-\left|S_m \right|}{F}$  and  $R_2=\max_{k_1 \in [K_1]}\left\{ \frac{|S_{k_1}|}{F}\right\}$.
\end{thm}
\begin{algorithm}
	\renewcommand{\thealgorithm}{2}
	\caption{Caching scheme based on $(K_1,K_2;F;Z_1,Z_2;S_m,S_1,..,S_{K_1})$ HPDA $\mathbf{Q}$ \cite{KYWM}}
	\label{alg2}
	\begin{algorithmic}[1]
		\Procedure{Placement}{$\mathbf{Q},\mathcal{W}$}       
		\State Split each file $W_n \in \mathcal{W}$ into $F$ packets, i.e., $W_n =\{W_{n,j} | j \in [F]\}$
		\For{\texttt{$k_1 \in [K_1]$}}
		\State  $\mathcal{Z}_{k_1}$ $\leftarrow$ $\{W_{n,j}: q_{j,k_1}^{(0)}=\star, n \in [N],j \in [F]\}$
		\EndFor
		\For{\texttt{$(k_1,k_2), k_1 \in [K_1], k_2 \in [K_2] $}}
		\State $\mathcal{Z}_{(k_1,k_2)}$ $\leftarrow$ $\{W_{n,j}: q_{j,k_2}^{(k_1)}=\star, n \in [N],j \in [F]\}$
		\EndFor
		\EndProcedure		
		\Procedure{Delivery Server}{$\mathbf{Q},\mathcal{W},\mathbf{d}$} 
		\For{\texttt{$s \in \left( \underset{k_1=1}{\bigcup^{K_1} S_{k_1}}\right) \textbackslash S_m $}}
		\State Server sends the following coded signal to the mirror sites:
		\State $X_s=\underset{q_{j,k_2}^{(k_1)}=s, j\in [F],k_1\in[K_1], k_2\in[K_2]}{\bigoplus}W_{d_{k_1,k_2},j}$
		\EndFor    
		\EndProcedure
		\Procedure{Delivery Mirrors}{$\mathbf{Q},\mathcal{W},\mathbf{d}, X_s $} 
		\For{\texttt{$k_1\in[K_1], s \in S_{k_1} \textbackslash S_m $}}
		\State After receiving $X_s$, mirror site $k_1$ sends the following coded signal to users in $\mathcal{U}_{k_1}$ :
		\State $X_{k_1,s}=X_s{\bigoplus}\left(\tiny \underset{\begin{array}{c} q_{j,k_2}^{(k'_1)}=s, q_{j,k_1}^{(0)}=\star, j\in [F], \\ k_2\in[K_2], k'_1\in[K_1]\textbackslash {k_1} \end{array}}{\bigoplus}W_{d_{k'_1,k_2},j}\right)$
		\EndFor
		 \For{\texttt{$k_1\in[K_1], s' \in S_{k_1} \bigcap S_m$}}
		 \State Mirror site $k_1$ sends the following coded signal to users in $\mathcal{U}_{k_1}$:
		 \State $X_{k_1,s'}=\underset{q_{j,k_2}^{(k_1)}=s', j\in [F],k_2\in[K_2]}{\bigoplus}W_{d_{k_1,k_2},j}$
		 \EndFor   
		\EndProcedure	
	\end{algorithmic}
\end{algorithm}

\section{Novel PDA constructions from $t$-designs}
In  this section, we construct two classes of PDAs based on $t$-designs.
\begin{figure*}[ht]
	\centering
	\begin{subfigure}[b]{\textwidth}
		\centering
		\captionsetup{justification=centering}
		\includegraphics[width=\textwidth]{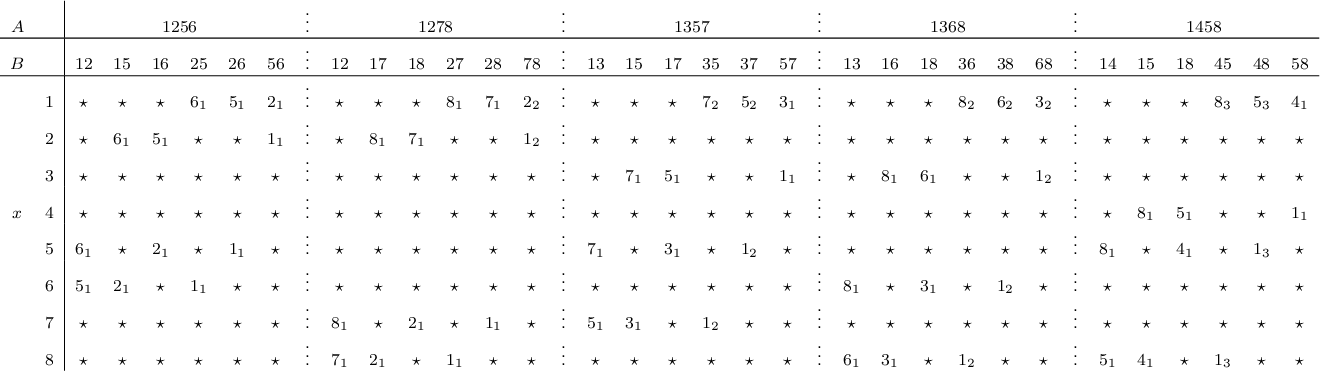}
		\caption{Column $1-30$ }
		\label{fig:example1_1}
	\end{subfigure}
	\hfill
	\begin{subfigure}[b]{\textwidth}
		\centering
		\captionsetup{justification=centering}
		\includegraphics[width=\textwidth]{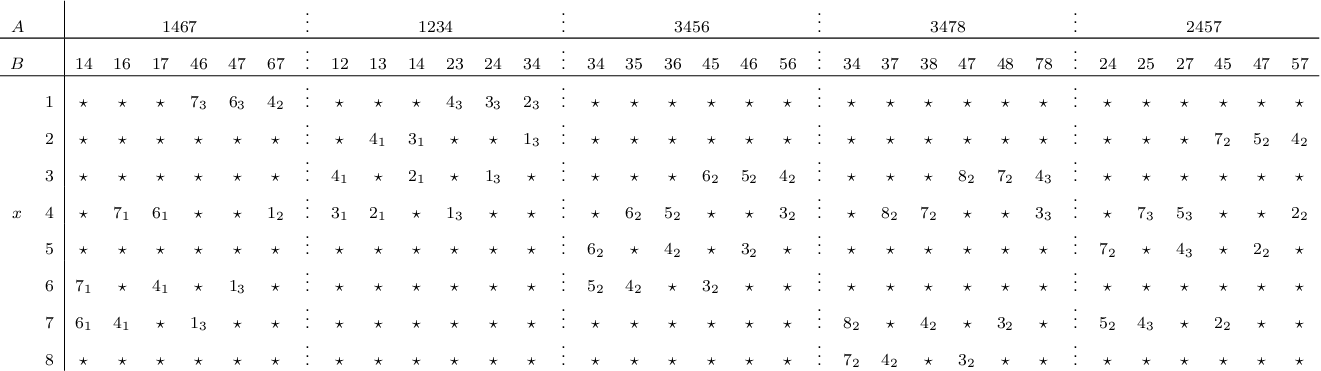}
		\caption{Column $31-60$ }
		\label{fig:example1_2}
	\end{subfigure}
	\hfill
	\begin{subfigure}[b]{\textwidth}
		\centering
		\captionsetup{justification=centering}
		\includegraphics[width=0.8\textwidth, keepaspectratio]{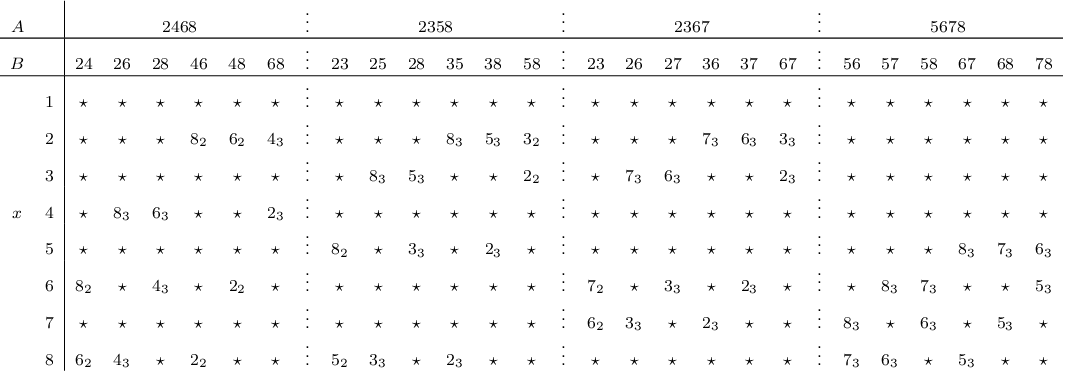}
		\caption{Column $61-84$ }
		\label{fig:example1_3}
	\end{subfigure}
	\caption{$8\times84$ array obtained from $3-(8,4,1)$ design for $i=2$ }
		\label{fig:example1}
\end{figure*}
\begin{figure*}[ht]
	\centering
	\begin{subfigure}[b]{\textwidth}
		\centering
		\captionsetup{justification=centering}
		\includegraphics[width=\textwidth]{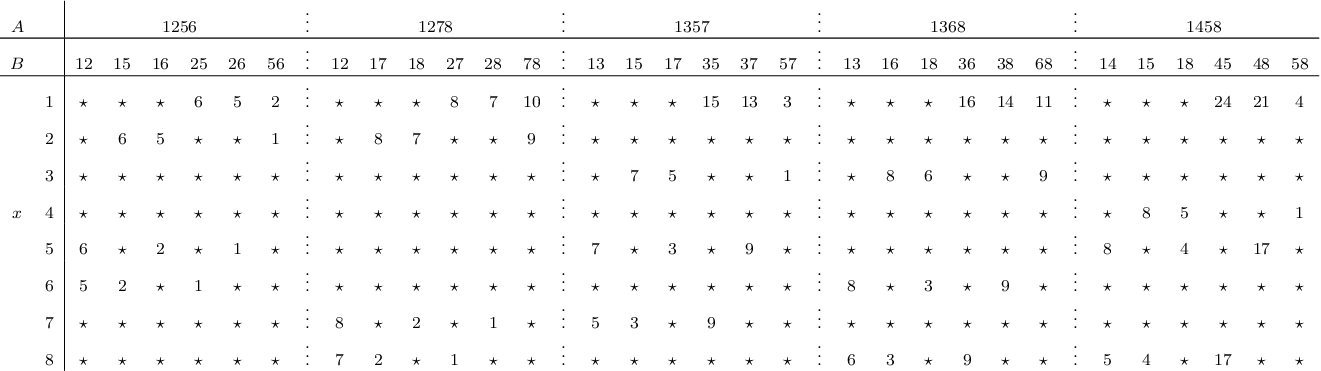}
		\caption{Column $1-30$ }
		\label{fig:example1_1_int}
	\end{subfigure}
	\hfill
	\begin{subfigure}[b]{\textwidth}
		\centering
		\captionsetup{justification=centering}
		\includegraphics[width=\textwidth]{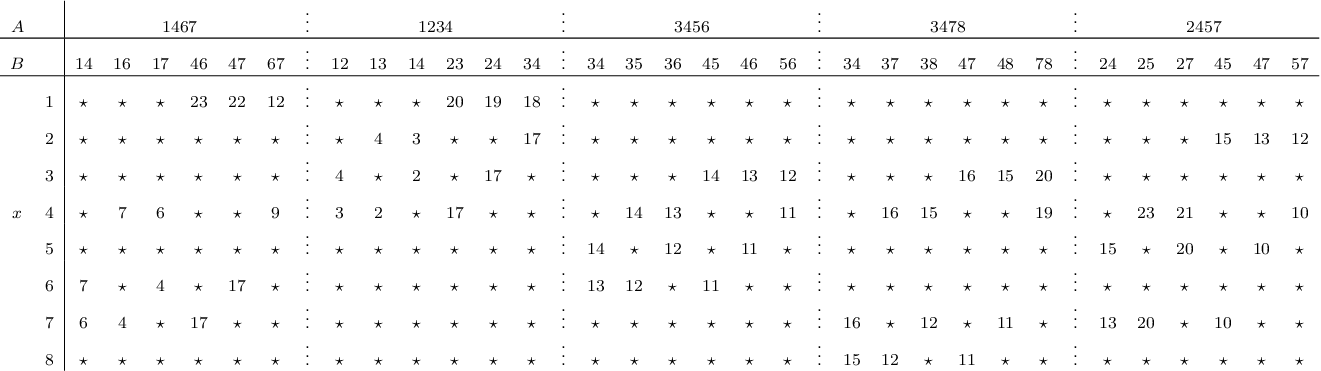}
		\caption{Column $31-60$ }
		\label{fig:example1_2_int}
	\end{subfigure}
	\hfill
	\begin{subfigure}[b]{\textwidth}
		\centering
		\captionsetup{justification=centering}
		\includegraphics[width=0.8\textwidth, keepaspectratio]{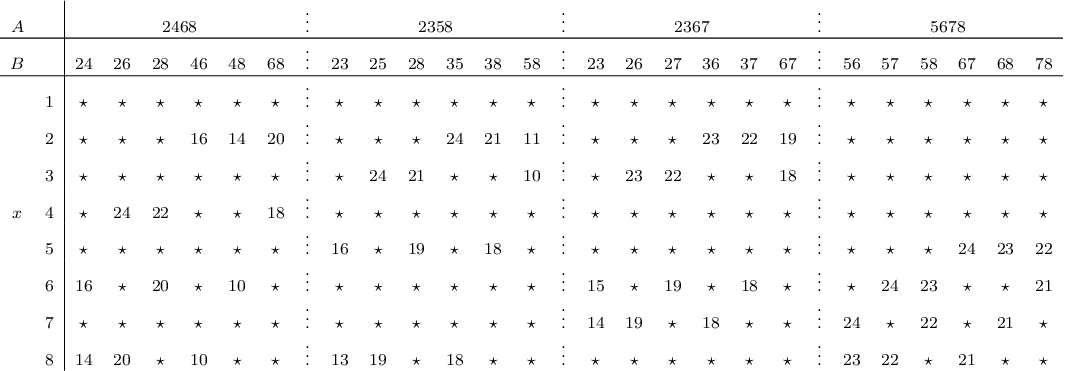}
		\caption{Column $61-84$ }
		\label{fig:example1_3_int}
	\end{subfigure}
	\caption{$(84,8,6,24)$ PDA obtained from $3-(8,4,1)$ design for $i=2$ }
	\label{fig:example1_int}
\end{figure*}
\subsection{Scheme I} We will refer to the PDA given in the following theorem by Scheme I.
	\begin{thm}\label{thm:scheme1}
		Given a $t-(v,k,\lambda)$ design, for any $i\in[1:t]$, there exists a $(v-i+1)-\left( \frac{\lambda \binom{v}{t}\binom{k}{i}}{\binom{k}{t}},v,v-i,\frac{\lambda\binom{v}{i-1}\binom{v-i}{t-i}}{\binom{k-i}{t-i}} \right)$ PDA which gives a coded caching scheme with memory ratio $\frac{M}{N}=1-\frac{i}{v}$ and rate $R=\frac{\lambda\binom{v}{i-1}\binom{v-i}{t-i}}{\binom{k-i}{t-i}v}$. This PDA can be written succinctly in terms of the parameters of the design as a $(v-i+1)-\left( b \binom{k}{i},v,v-i,\binom{v}{i-1} \lambda_i \right)$ PDA.
	\end{thm}
	\begin{IEEEproof}
	Let $(\mathcal{X}, \mathcal{A})$ be a $t-(v,k,\lambda)$ design. An array $\mathbf{P}$  whose rows are indexed by all the points of $\mathcal{X}$ and columns indexed by all the elements of $C=\{ (A,B) : A \in \mathcal{A}, B \in \binom{A}{k-i} \}$ for any $i \in [1:t]$, is defined as follows.	
	\begin{equation}\label{eq:cons1}
		P_{x,(A,B)}= \begin{cases}  (A\textbackslash\{B\cup x\})_\alpha, & \text {if } x \in A \text{ and } x \notin B    \\ \hspace{0.5 cm} \star & \text {otherwise }\end{cases},
	\end{equation}
where the subscript $\alpha$ denotes the $\alpha^{th}$ occurrence of $A\textbackslash\{B\cup x\}$ from left to right in the row indexed by $x$.
	The subpacketization level $F$ is the number of rows of $\mathbf{P}$, which is equal to the number of points in  $\mathcal{X} = v$. The number of users $K$ is given by the number of columns of $\mathbf{P}$. Since columns are indexed by $C=\{ (A,B): A \in \mathcal{A}, B \in \binom{A}{k-i} \}$, the number of columns is the number of blocks in $\mathcal{A}$ times the number of $(k-i)$ sized subsets of $k$ sized block. Therefore,  $K=\frac{\lambda\binom{v}{t}}{\binom{k}{t}}\binom{k}{k-i}=\frac{\lambda\binom{v}{t}}{\binom{k}{t}}\binom{k}{i}$.
	
The non $\star$ entries of $\mathbf{P}$ are denoted by $(A\textbackslash\{B\cup x\})_\alpha$. Thus, by $(\ref{eq:cons1})$, $|A\textbackslash\{B\cup x\}|=k-(k-i+1)=(i-1)$. Therefore, the number of possible $A\textbackslash\{B\cup x\}$ is $\binom{v}{i-1}$. To find the number of occurrences of $A\textbackslash\{B\cup x\}$ in a given row $x \in \mathcal{X}$, we have to find the number of blocks in which $A\textbackslash\{B\cup x\}$ and $x$ belong. From $(\ref{eq:cons1})$, $|A\textbackslash\{B\cup x\} \cup x|=i$. Therefore, by Theorem~\ref{thm:Stin} and since $i \in [1,t]$, the number of blocks in which $i$ sized subset of $\mathcal{A}$ belong is equal to $\frac{\lambda\binom{v-i}{t-i}}{\binom{k-i}{t-i}}$. Therefore, 
\begin{equation*}
	\alpha \in \left[1,\frac{\lambda\binom{v-i}{t-i}}{\binom{k-i}{t-i}}\right]. \\
	\text{ Thus } S= \binom{v}{i-1}\frac{\lambda\binom{v-i}{t-i}}{\binom{k-i}{t-i}}.
\end{equation*}
\begin{figure*}[ht]
	\begin{center}
		\captionsetup{justification=centering}
		\includegraphics[width=0.9\textwidth]{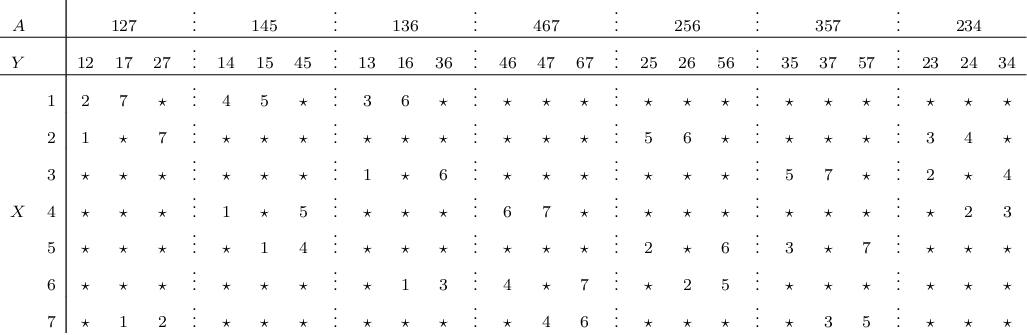}
		\caption{ $(21,7,5,7)$ PDA}
		\label{fig:example2}
	\end{center}
\end{figure*}
Let $\mathcal{S}$ denote all the $(i-1)$ sized subsets of $\mathcal{X}$  arranged in some order, then we define a bijection $f(.)$ from $\left\{\mathcal{S} \subset \mathcal{X} : |\mathcal{S}|=(i-1)\right\}$ to $\left[1,\binom{v}{i-1} \right]$. Thus, we can replace the non $\star$ entries of $\mathbf{P}$,	$(A\textbackslash\{B\cup x\})_\alpha$, by $(\alpha-1)\binom{v}{i-1}+f\left(A\textbackslash\{B\cup x\}\right)$. This gives an array $\mathbf{P}$ with $S$ distinct non-negative integers. Now we have to show that the array $\mathbf{P}$ is the PDA specified in Theorem \ref{thm:scheme1}, which is given in \textit{Appendix B}.
\end{IEEEproof}
  \begin{example}
	$\mathcal{X}=\{1,2,3,4,5,6,7,8\}, \\ \mathcal{A}=\{1256,1278,1357,1368, 1458,1467,1234,3456,3478, \\ 2457, 2468,2358,2367,5678\}$ is a $3-(8,4,1)$ design.	From this design for $i=2$, (\ref{eq:cons1}) gives a $8\times84$ array shown in \autoref{fig:example1}. A mapping from $a_{\alpha}$ to $(\alpha-1)8+a$, where $a\in[1:8], \alpha\in[1:3]$, will give a $(84,8,6,24)$ PDA shown in \autoref{fig:example1_int}.
\end{example}

\subsection{Scheme II} We will refer to the PDA given in the following theorem by Scheme II.
\begin{thm}\label{thm:scheme2}
	Given a $t-(v,k,\lambda)$ design, for any $i \in [1:t-1]$, there exists a $\binom{v-1}{i}-\left( \frac{\lambda\binom{v}{t}\binom{k}{i+1}}{\binom{k}{t}},\binom{v}{i},\binom{v}{i}-(i+1),\frac{v\lambda\binom{v-(i+1)}{t-(i+1)}}{\binom{k-(i+1)}{t-(i+1)}} \right)$ PDA which gives a coded caching scheme with memory ratio $\frac{M}{N}=1-\frac{i+1}{\binom{v}{i}}$ and rate $R=\frac{v\lambda\binom{v-(i+1)}{t-(i+1)}}{\binom{k-(i+1)}{t-(i+1)}\binom{v}{i}}$. This PDA can be written succinctly in terms of the parameters of the design as a  $\binom{v-1}{i}-\left( b\binom{k}{i+1},\binom{v}{i},\binom{v}{i}-(i+1),v \lambda_{i+1} \right)$ PDA.
\end{thm}
\begin{IEEEproof}
	Let $(\mathcal{X}, \mathcal{A})$ be a $t-(v,k,\lambda)$ design. For any $i \in [1:t-1]$, define an array $\mathbf{P}$  whose rows are indexed by $X \in \binom{\mathcal{X}}{i}$  and columns indexed by all the elements of $C=\{ (A,Y) : A \in \mathcal{A}, Y \in \binom{A}{i+1}\}$, as follows.
	
	\begin{equation}\label{eq:cons2}
		P_{X,(A,Y)}= \begin{cases}  (Y\textbackslash X)_\alpha, & \text {if } X \subset Y   \\ \hspace{0.5 cm} \star & \text {otherwise }\end{cases},
	\end{equation}
where the subscript $\alpha$ denotes the $\alpha^{th}$ occurrence of $Y\textbackslash X$ from left to right in the row indexed by $X$. The subpacketization level $F$ is the number of rows of $\mathbf{P}$, which is equal to $\binom{v}{i}$. The number of users $K$ is given by the number of columns of $\mathbf{P}$. Since columns are indexed by $C=\{ (A,Y) : A \in \mathcal{A}, Y=\binom{A}{i+1}\}$, the number of columns is the number of blocks in $\mathcal{A}$ times the number of $(i+1)$ sized subsets of $k$ sized blocks. Therefore,  $K=\frac{\lambda\binom{v}{t}}{\binom{k}{t}}\binom{k}{i+1}$. 

The non $\star$ entries of $\mathbf{P}$ are denoted by $(Y\textbackslash X)_\alpha$ and $Y\textbackslash X \in \mathcal{X} $. Therefore, the number of possible $Y\textbackslash X$ is $v$. To find the number of occurrences of $Y\textbackslash X$ in a given row $X$, by $(\ref{eq:cons2})$, we need to find the number of blocks in which $\{Y\textbackslash X\} \cup X = Y$ belongs. By Theorem~\ref{thm:Stin}, that is equal to $\frac{\lambda\binom{v-(i+1)}{t-(i+1)}}{\binom{k-(i+1)}{t-(i+1)}}$. Therefore, 
\begin{equation*}
	\alpha \in \left[1,\frac{\lambda\binom{v-(i+1)}{t-(i+1)}}{\binom{k-(i+1)}{t-(i+1)}}\right]. \\
	\text{ Thus } S= \frac{v\lambda\binom{v-(i+1)}{t-(i+1)}}{\binom{k-(i+1)}{t-(i+1)}}.
\end{equation*}

Let $x$ denote all the points of $\mathcal{X}$  arranged in some order, then we define a bijection $f(.)$ from $\left\{x \in \mathcal{X} \right\}$ to $[1, v]$. Thus, we can replace all the non $\star$ entries of $\mathbf{P}$,	$(Y\textbackslash X)_\alpha$, by $(\alpha-1)v+f\left(Y\textbackslash X\right)$. This gives an array $\mathbf{P}$ with $S$ distinct non-negative integers. Now we have to show that the array $\mathbf{P}$ is the PDA specified in Theorem \ref{thm:scheme2}, which is given in \textit{Appendix C}.
\end{IEEEproof}
  \begin{example}
  	Consider the $2-(7,3,1)$ design in Example $\ref{ex:example1}$. For $i=1$, we get the $(21,7,5,7)$ PDA shown in \autoref{fig:example2}. 
  \end{example}

\section{Performance Analysis}
In this section, we compare the performance of the proposed schemes with existing coded caching schemes from $t$-designs. The optimality of the obtained PDAs is also discussed. 
\subsection{Scheme I }
In \cite{Li}, a PDA construction from a simple $t-(v,k,\lambda)$ design with $k \leq 2t$ is given. Compared with that for $\lambda=1$, the proposed scheme has improved rate in the cases where the parameters $K, F \text{ and } \frac{M}{N}$ are the same.
\begin{example}
	For a $2-(7,3,1)$ design in Example \ref{ex:example1} and for $i=1$, the proposed Scheme I gives a $(21,7,6,3)$ PDA giving a coded caching scheme with $\frac{M}{N}=\frac{6}{7}$ and rate $R=0.4285$, whereas Theorem $6$ in \cite{Li} gives a $ (21,7,6,7)$ PDA giving a coded caching scheme with same memory ratio and rate $R=1$. 
\end{example}
\begin{table*}[ht]
	\centering
	\caption{Comparison of our Scheme II with known constructions from $3-(8,4,1)$ design}
	\begin{tabular}{| c | c | c | c | c| c|}
		\hline
		\rule{0pt}{4ex}
		\makecell{Schemes and parameters} & \makecell{Number of users\\ $K$} & \makecell{Caching ratio \\ $\frac{M}{N}=\frac{Z}{F}$} & \makecell{Subpacketization \\ $F$} & \makecell{Number of integers\\ $S$}& \makecell{Rate\\ $R=\frac{S}{F}$}  \\ [4pt] 		
		\hline
		\rule{0pt}{3.5ex}
		\makecell{Proposed scheme II  \\  for $i=1$} & $84$ & $\frac{6}{8}$ & $8$ & $24$ & $3$ \\
		\hline
		\rule{0pt}{3.5ex}
		\makecell{Theorem 5 in \cite{Li} \\  for $i=2$ }  & $84$ & $\frac{6}{8}$ & $8$ & $56$ & $7$ \\		
		\hline
		\rule{0pt}{3.5ex}
		\makecell{Proposed scheme II  \\  for $i=2$} & $56$ & $\frac{25}{28}$ & $28$ & $8$ & $0.2857$\\
		\hline
		\rule{0pt}{3.5ex}
		\makecell{Theorem 5 in \cite{Li} \\  for $i=1$ }  & $56$ & $\frac{25}{28}$ & $28$ & $24$ & $0.8571$ \\
		\hline
		\rule{0pt}{3.5ex}
		\makecell{Proposed scheme II transpose  \\  for $i=2$ }  & $28$ & $\frac{25}{28}$ & $56$ & $8$ & $0.1428$ \\
		\hline
		\rule{0pt}{3.5ex}
		\makecell{Scheme I in \cite{SSP}  \\    }  & $28$ & $\frac{25}{28}$ & $56$ & $24$ & $0.4285$ \\
		\hline
		\rule{0pt}{3.5ex}
		\makecell{ MN Scheme in \cite{MaN}  \\  parameter $K=28$ and $t=\frac{KM}{N}=25$  }  & $28$ & $\frac{25}{28}$ & $3276$ & $378$ & $0.1153$ \\
		\hline
	\end{tabular}
	\label{tab:comp}
\end{table*}
\begin{thm}\label{thm:opt_schm_1}
	The class of PDAs obtained in \textit{Theorem \ref{thm:scheme1}} is a class of \textit{optimal PDAs}. \text{(Proof given in \textit{Appendix D}.)} 
\end{thm}
\begin{rem}
In Scheme I, for $\lambda=1$ and $i=t$, we get a $\left( \binom{v}{t},v,v-t,\binom{v}{t-1} \right)$ PDA. This is indeed the first variant MN PDA given in \cite{MJXQ}. This is also an RPDA given in Theorem $3.15$ in \cite{Wei} which is constructed recursively. It can be verified from $(\ref{eq:cons1})$ that the number of $\star$ is same in all rows of $\mathbf{P}$. Therefore, transpose of this PDA gives a $\left(v,\binom{v}{t},\binom{v}{t}(1-\frac{t}{v}),\binom{v}{t-1} \right)$ PDA which gives a coded caching scheme with $\left(K=v, F=\binom{v}{t}, \frac{M}{N}=(1-\frac{t}{v}) \text{ and } R= \frac{t}{v-t+1} \right)$. \\ This specific case of scheme I is same as the scheme II construction in \cite{SSParx} from $t-(v, k, 1)$ design. This PDA indeed is the PDA corresponding to the MN scheme in \cite{MaN} for the same parameters. 
\end{rem}

\subsection{Scheme II}

The proposed construction in Scheme II has improved transmission rate compared to existing schemes from $t-(v, k, 1)$ designs in \cite{SSP} and in \cite{Li} for the same number of users, subpacketization and memory ratio. 

For a $t-(v, k, 1)$ design with $k=t+1$, the proposed Scheme II in Theorem $6$ gives a $\left( \frac{\binom{v}{t}\binom{t+1}{i+1}}{t+1},\binom{v}{i},\binom{v}{i}-(i+1),\frac{v\binom{v-(i+1)}{t-(i+1)}}{t-i} \right)$ PDA for any $i \in [1:t-1]$. The PDA construction proposed in Theorem $5$ in \cite{Li} from $t-(v, k, 1)$ design for any $i \in [1:t-1]$ have the same $K, F \text{ and } \frac{M}{N}$ as in the proposed Scheme II, but a larger value of $S$ (Comparing the proposed scheme for $i$ and scheme in \cite{Li} for $t-i$). Thus Scheme II has improved $R$ for the same set of parameters. This is illustrated with an example of $3-(8,4,1)$ design in the Table \ref{tab:comp}. 

From $(\ref{eq:cons2})$, it can be seen that the number of $\star$ in any row of $\mathbf{P}$, $ Z'= \text{number of columns } - \text{number of non }\star \text{ elements in a row } = \frac{\lambda\binom{v}{t}\binom{k}{i+1}}{\binom{k}{t}} - \frac{\lambda\binom{v-i}{t-i}(k-i)}{\binom{k-i}{t-i}}= \frac{\lambda \binom{v}{t}\binom{k}{i+1}}{\binom{k}{t}} \left( 1-\frac{i+1}{\binom{v}{i}}\right) $. Therefore, the transpose of $\mathbf{P}$ also gives a PDA. For a $t-(v, k, 1)$ design with $k=t+1$ and $i=t-1$, the transpose of PDA given in Scheme II gives a $\left( \binom{v}{t-1},\binom{v}{t},\binom{v}{t}\left(1-\frac{t}{\binom{v}{t-1}}\right),v \right)$ PDA which corresponds to a coded caching scheme with memory ratio $\frac{M}{N}=\left(1-\frac{t}{\binom{v}{t-1}}\right)=\left(1-\frac{v-t+1}{\binom{v}{t}}\right)$ and rate $R=\frac{v}{\binom{v}{t}}$. The Scheme I proposed in \cite{SSP} has the same $K,F \text{ and } \frac{M}{N}$ as in the proposed scheme for a $t-(v, k, 1)$ design with $k=t+1$, but the transmission rate of the proposed scheme is improved by a factor of $t$. This is illustrated with an example of $3-(8,4,1)$ design in the Table \ref{tab:comp}. A comparison with the MN scheme with same $K\text{ and } \frac{M}{N}$, shows that with a small increase in rate, the subpacketization is reduced significantly. 
\begin{rem}
For $\lambda=1$ and $i=t-1$, we obtain a $\left(\binom{v}{t},\binom{v}{t-1},\binom{v}{t-1}-t,v\right)$ PDA which is indeed a case of the Second Variant MN PDA given in \cite{MJXQ}.
\end{rem}
\begin{thm}\label{thm:opt_schm_2}
	The PDA obtained in \textit{Theorem \ref{thm:scheme2}} is an \textit{optimal PDA} for $i=1$. For $i=(t-1)$, the PDA obtained and the transposed PDA is optimal when $\lambda=1$.
\end{thm}
(Proof given in \textit{Appendix E}.)
\begin{figure*}[ht]
	\centering
	\begin{subfigure}[b]{\textwidth}
		\centering
		\captionsetup{justification=centering}
		\includegraphics[width=\textwidth]{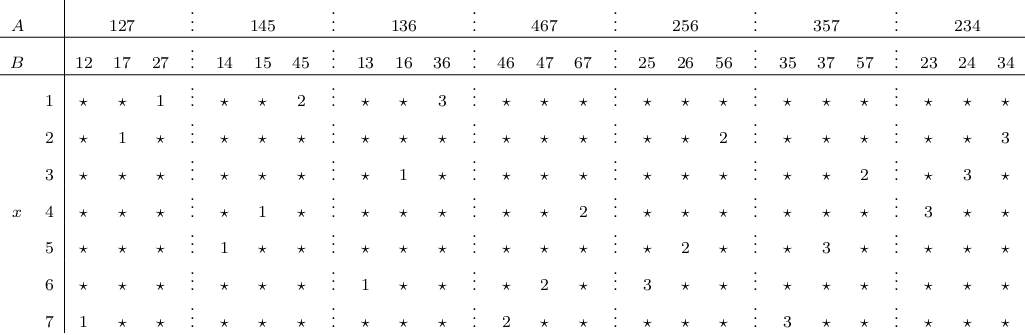}
		\caption{$(21,7,6,3)$ PDA constructed by Scheme I using $2-(7,3,1)$ design for $i=1$}
		\label{fig:hpda1_1}
	\end{subfigure}
	\hfill
	\begin{subfigure}[b]{\textwidth}
		\centering
		\captionsetup{justification=centering}
		\includegraphics[width=\textwidth]{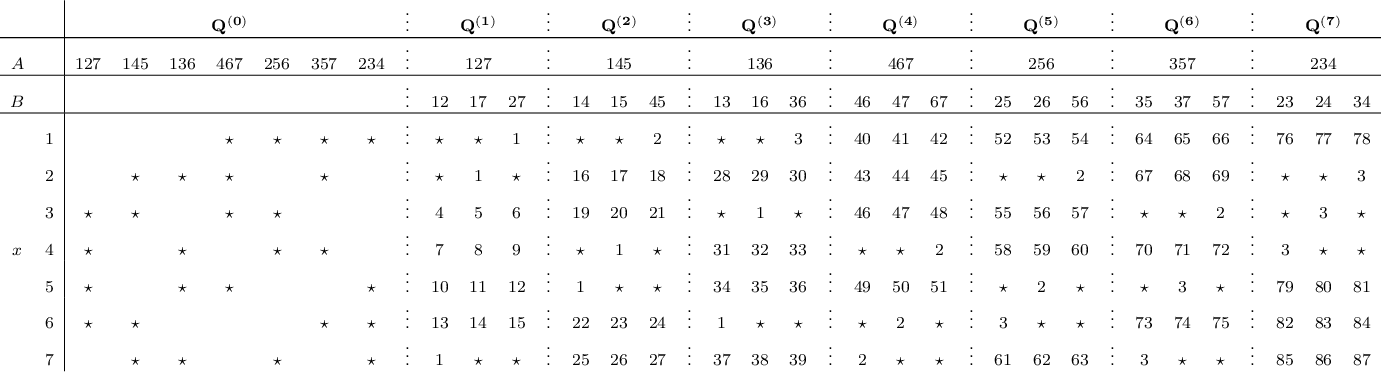}
		\caption{ HPDA from the $(21,7,6,3)$ PDA in \autoref{fig:hpda1_1} }
		\label{fig:HPDA1}
	\end{subfigure}
	\caption{Example of HPDA in Theorem \ref{thm:HPDA}}
\end{figure*}
\begin{rem}
	Checking the optimality of constructed PDAs in Scheme II for $i \in [2,t-2] $ is hard in general. As an example,  for a $4-(23,7,1)$ design for $i=2$, we get a $(8855,253,250,115)$ PDA with rate $R=0.4545$. For this PDA, by  ($\ref{eq:MJXQ}$), $S \geq 105 + 1+1= 107$. Since Scheme II construction gives a \textit{g-PDA}, $S$ must be a factor of $K(F-Z)=26565$. The smallest factor of $26565$ which satisfies the bound $S \geq 107$ is $115$, which is the $S$ obtained. Therefore, we can say, this is \textit{an optimal regular PDA}. A non-regular PDA with $ 107 \leq S \le 115$ exists or not in this case is an open question. 
\end{rem}
\begin{rem}
	  For a $2-(7,3,1)$ design for $i=2$, \cite{MACC_arx} gives a $(7,63,45,35)$ PDA with memory ratio $\frac{5}{7}$ and rate $R=0.555$. For the same $2-(7,3,1)$ design for $i=1$, transpose of proposed scheme II gives a $(7,21,15,7)$ PDA with memory ratio $\frac{5}{7}$ and rate $R=0.333$. So, for the same $K=7$ and $\frac{M}{N}=\frac{5}{7}$, proposed scheme gives an improved rate with less subpacketization. However, a general comparison of the proposed scheme and the PDA construction in Theorem 1 of \cite{MACC_arx} is not possible as both give different parameters of $K, F, Z$ and $S$. 
\end{rem}
\section{HPDA from $t$-designs}
Hierarchical coded caching schemes with low subpacketization level can be achieved by constructing appropriate HPDAs.  
Two classes of PDAs proposed in Section III have good performance compared to existing coded caching schemes from $t$-designs and have shown to be optimal in many cases in Section IV. In this section, we construct HPDAs from the above proposed PDAs and analyze their performance.
\begin{thm}\label{thm:HPDA}
	Given a $t-(v,k,\lambda)$ design, for any $i \in [1:t]$, there exist a $\left(\frac{\lambda\binom{v}{t}}{\binom{k}{t}},\binom{k}{k-i};v;(v-k),(k-i);S_m,S_1,..,S_{K_1} \right)$ HPDA which gives an $v$ division $\left(\frac{\lambda\binom{v}{t}}{\binom{k}{t}},\binom{k}{k-i};M_1,M_2;N \right)$ coded caching scheme with memory ratios $\frac{M_1}{N}=1-\frac{k}{v}$, $\frac{M_2}{N}=\frac{k-i}{v}$  and transmission load $R_1=\frac{\lambda\binom{v}{i-1}\binom{v-i}{t-i}}{\binom{k-i}{t-i}v}$ and $R_2 \leq \frac{\binom{k}{i-1}\frac{\lambda\binom{v-i}{t-i}}{\binom{k-i}{t-i}}+(v-k)\binom{k}{k-i}}{v}$. \\
	 \text{(Proof given in \textit{Appendix F}.)}
\end{thm}
For $\lambda=1$ and $i=t$, Theorem \ref{thm:HPDA} gives a $\left(\frac{\binom{v}{t}}{\binom{k}{t}}, \binom{k}{k-t}; v; (v-k), (k-t); S_m, S_1, .., S_{K_1} \right)$ HPDA which gives an $v$ division $\left(\frac{\binom{v}{t}}{\binom{k}{t}},\binom{k}{k-t};M_1,M_2;N \right)$ coded caching scheme with memory ratios $\frac{M_1}{N}=1-\frac{k}{v}$, $\frac{M_2}{N}=\frac{k-t}{v}$  and transmission load $R_1=\frac{\binom{v}{t-1}}{v}$ and $R_2 = \frac{\binom{k}{t-1}+(v-k)\binom{k}{k-t}}{v}$.
\begin{example}
	From the $(21,7,6,3)$ PDA constructed by Scheme I using $2-(7,3,1)$ design for $i=1$ shown in \autoref{fig:hpda1_1}, we can obtain the HPDA shown in \autoref{fig:HPDA1}. 
\end{example}  
\begin{figure*}[ht]
	\begin{center}
		\captionsetup{justification=centering}
		\includegraphics[width=\textwidth]{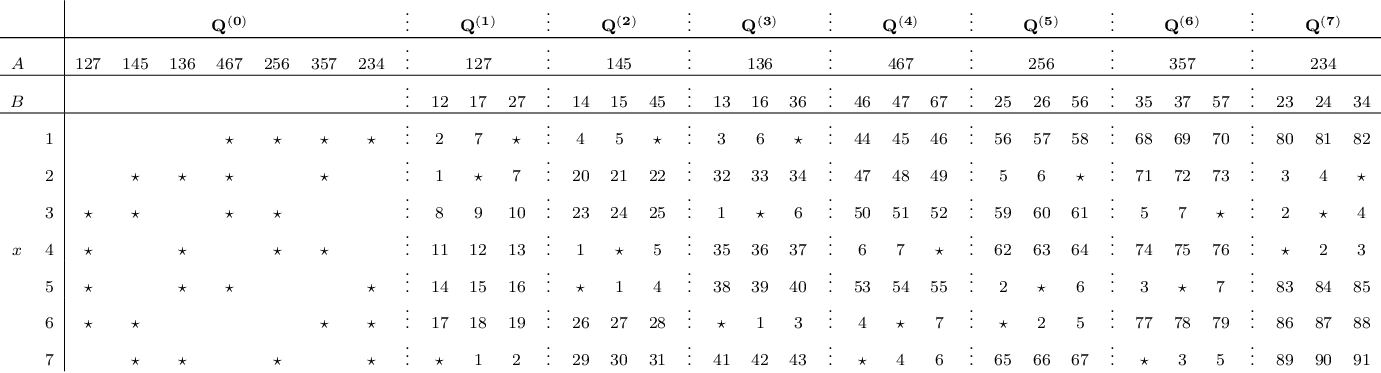}
		\caption{ Example of HPDA in Theorem \ref{thm:HPDA2}}
		\label{fig:HPDA2}
	\end{center}
\end{figure*}
\begin{thm}\label{thm:HPDA2}
	Given a $t-(v,k,\lambda)$ design, for any $i \in [1:t-1]$, there exist a $\left(\frac{\lambda\binom{v}{t}}{\binom{k}{t}},\binom{k}{i+1};\binom{v}{i};\binom{v}{i}-\binom{k}{i},\binom{k}{i}-(i+1);S_m,S_1,..,S_{K_1} \right)$ HPDA which gives an $\binom{v}{i}$ division $\left(\frac{\lambda\binom{v}{t}}{\binom{k}{t}},\binom{k}{i+1};M_1,M_2;N \right)$ coded caching scheme with memory ratios $\frac{M_1}{N}=1-\frac{\binom{k}{i}}{\binom{v}{i}}$, $\frac{M_2}{N}=\frac{\binom{k}{i}-(i+1)}{\binom{v}{i}}$  and transmission load $R_1=\frac{v\lambda\binom{v-(i+1)}{t-(i+1)}}{\binom{k-(i+1)}{t-(i+1)}\binom{v}{i}}$ and $R_2 \leq \frac{\frac{k\lambda\binom{v-(i+1)}{t-(i+1)}}{\binom{k-(i+1)}{t-(i+1)}}+\left[\binom{v}{i}-\binom{k}{i}\right]\binom{k}{i+1}}{\binom{v}{i}}$.
\end{thm}
Similar to the HPDA obtained in Theorem \ref{thm:HPDA} from PDA constructed in Scheme I, we can obtain HPDA in Theorem \ref{thm:HPDA2} from PDA constructed in Scheme II.
For $\lambda=1$ and $i=t-1$, Theorem \ref{thm:HPDA2} gives a $\left(\frac{\binom{v}{t}}{\binom{k}{t}},\binom{k}{t};\binom{v}{t-1};\binom{v}{t-1}-\binom{k}{t-1},\binom{k}{t-1}-t;S_m,S_1,..,S_{K_1} \right)$ HPDA which gives an $\binom{v}{t-1}$ division $\left(\frac{\binom{v}{t}}{\binom{k}{t}},\binom{k}{t};M_1,M_2;N \right)$ coded caching scheme with memory ratios $\frac{M_1}{N}=1-\frac{\binom{k}{t-1}}{\binom{v}{t-1}}$, $\frac{M_2}{N}=\frac{\binom{k}{t-1}-t}{\binom{v}{t-1}}$  and transmission load $R_1=\frac{v}{\binom{v}{t-1}}$ and $R_2 = \frac{k+\left[\binom{v}{t-1}-\binom{k}{t-1}\right]\binom{k}{t}}{\binom{v}{t-1}}$.
\begin{example}
	From the $(21,7,5,7)$ PDA constructed by Scheme II using $2-(7,3,1)$ design for $i=1$ shown in \autoref{fig:example2}, we can obtain the HPDA shown in \autoref{fig:HPDA2}. 
\end{example} 
\begin{table*}[hb]
	\centering
	\caption{Comparison of hierarchical coded caching schemes}
	\begin{tabular}{| c | c | c | c | c | c | c | c | c |}
		\hline
		\rule{0pt}{4ex}
		\makecell{Schemes and parameters}  & \makecell{$K_1$} & \makecell{$K_2$}& \makecell{$\frac{M_1}{N}$} & \makecell{$\frac{M_2}{N}$} & \makecell{$F$}& \makecell{$R_1$} &\makecell{$R_2$}& \makecell{$T=R_1+R_2$} \\ [4pt] 	
		\hline
		\rule{0pt}{3.5ex}
		\makecell{Proposed scheme in Theorem \ref{thm:HPDA}  \\  for $3-(8,4,1)$ design with $i=3$} & $14$ & $4$& $\frac{1}{2}$ & $\frac{1}{8}$ & $8$ & $3.5$ & $2.75$ & $6.25$ \\	
		\hline
		\rule{0pt}{3.5ex}
		\makecell{Scheme in \cite{KNMD} with centralized \\ data placement ($\alpha^{\star}=\frac{1}{2}, \beta^{\star}=\frac{1}{4}$)} & $14$ & $4$& $\frac{1}{2}$ & $\frac{1}{8}$ & $1.489\times10^{11}$ & $1.98295$ & $2.75$ & $4.73295$ \\
		\hline
		\rule{0pt}{3.5ex}
		\makecell{Proposed scheme in Theorem \ref{thm:HPDA}  \\  for $3-(8,4,1)$ design with $i=2$} & $14$ & $6$ & $\frac{1}{2}$& $\frac{1}{4}$ & $8$ & $3$ & $4$ & $7$ \\
		\hline
		\rule{0pt}{3.5ex}
		\makecell{Scheme in \cite{KNMD} with centralized \\ data placement ($\alpha^{\star}=\frac{1}{2}, \beta^{\star}=\frac{1}{4}$)}  & $14$ & $6$ & $\frac{1}{2}$& $\frac{1}{4}$ & $1.56\times10^{23}$ & $0.808$ & $2.28125$ & $3.08925$ \\
		\hline
		\rule{0pt}{3.5ex}
		\makecell{Scheme in Theorem 3 \\ in \cite{KYWM} with $\mathbf{A}$ and $\mathbf{B}$ are the MN PDA \\ and an optimal PDA proposed in \cite{Wei} respectively. \\ $\mathbf{A}=(14, 3432,1716,3003)$ PDA \\ $\mathbf{B}=(6,4,1,11)$ PDA  given by Theorem 4.2 in \cite{Wei}}  & $14$ & $6$ & $\frac{1}{2}$& $\frac{1}{4}$ & $1.37\times10^{4}$ & $2.406$ & $2.75$ & $5.156$ \\
		\hline
		\rule{0pt}{3.5ex}
		\makecell{Scheme in Theorem 3 \\ in \cite{KYWM} with $\mathbf{A}$ and $\mathbf{B}$ are the optimal PDAs \\ given by Theorem 4.2 in \cite{Wei}. $\mathbf{A}=(14,2,1,7)$ PDA \\ $\mathbf{B}=(6,4,1,11)$ PDA }  & $14$ & $6$ & $\frac{1}{2}$& $\frac{1}{4}$ & $8$ & $9.625$ & $2.75$ & $12.375$ \\
		\hline
		\rule{0pt}{3.5ex}
		\makecell{Proposed scheme in Theorem \ref{thm:HPDA2}  \\  for $3-(8,4,1)$ design with $i=2$} & $14$ & $4$ & $\frac{11}{14}$& $\frac{3}{28}$ & $28$ & $0.2857$ & $3.285$ & $3.5714$ \\
		\hline
\rule{0pt}{3.5ex}
		\makecell{Scheme in \cite{KNMD} with centralized \\ data placement ($\alpha^{\star}=\frac{11}{14}, \beta^{\star}=\frac{1}{4}$)}  & $14$ & $4$ & $\frac{11}{14}$& $\frac{3}{28}$ & $1.346\times10^{15}$ & $0.3409$ & $3.107$ & $3.448$ \\
		\hline
\rule{0pt}{3.5ex}
		\makecell{Proposed scheme in Theorem \ref{thm:HPDA2}  \\  for $3-(26,6,1)$ design with $i=2$} & $130$ & $20$ & $\frac{62}{65}$& $\frac{12}{325}$ & $325$ & $0.08$ & $19.095$ & $19.1753$ \\
		\hline
\rule{0pt}{3.5ex}
		\makecell{Scheme in \cite{KNMD} with centralized \\ data placement ($\alpha^{\star}=\frac{62}{65}, \beta^{\star}=\frac{1}{4}$)}  & $130$ & $20$ & $\frac{62}{65}$& $\frac{12}{325}$ & $1.39\times10^{1824}$ & $0.0307$ & $17.1668$ & $17.1975$ \\
		\hline
	\end{tabular}
	\label{tab:comp_HPDA}
\end{table*}
\subsection{Performance analysis}
To compare the performance of different hierarchical coded caching schemes, the metric considered in \cite{WWCY} is \textit{coding delay}, $T$, defined as the duration of the delivery phase normalized to the file size. If parallel transmission between the two layers is possible, then $T=\max\{R_1,R_2\}$. On the other hand, if mirror transmission occurs after server finishes the transmission, then $T=R_1+R_2$ \cite{WWCY}. The goal is to design a hierarchical caching scheme that minimizes the coding delay \cite{WWCY}. 

Under uncoded placement, the hierarchical coded caching schemes in \cite{KNMD} and \cite{WWCY} achieves optimal $R_2$ when $\alpha=\beta$. The scheme in Theorem 2 in \cite{KYWM} achieves optimal $R_1$ under uncoded placement. But, the schemes in \cite{KNMD}, \cite{WWCY} and Theorem 2 in \cite{KYWM} requires a subpacketization level that grows exponentially with the number of total users. Thus for practical implementations, it is desirable to have hierarchical schemes with reduced subpacketization levels. A hierarchical scheme using HPDA construction from two PDAs with low subpacketization level is given in Theorem 3 in \cite{KYWM}.  Our proposed hierarchical schemes in Theorem \ref{thm:HPDA} and Theorem \ref{thm:HPDA2} have low subpacketization levels. A comparison of proposed schemes and scheme in Theorem 3 in \cite{KYWM} is hard in general.

A comparison of proposed hierarchical schemes and the existing schemes is given in Table \ref{tab:comp_HPDA}. The subpacketization level of schemes in \cite{KNMD} is taken as the maximum subpacketization level required while memory sharing.  The comparison shows that the proposed class of HPDAs in Theorem \ref{thm:HPDA} gives schemes with low subpacetization levels at the expense of increase in coding delay. A scheme obtained through construction in Theorem 3 in \cite{KYWM} using two optimal PDAs which achieves the same subpacketization as ours has larger coding delay than ours. The proposed class of HPDAs in Theorem \ref{thm:HPDA2} gives schemes with significant reduction in subpacketization level at the expense of a small increase in the coding delay. 
In general, the proposed HPDA constructions from $t$-designs give low subpacketization level schemes with a larger mirror cache size and a low user cache size.

\section{Conclusion}
In this work, we introduced two novel classes of PDA constructions from $t-(v,k,\lambda)$ designs. Scheme I achieves an optimal rate for a given $K, F$ and $Z$. Scheme II achieves an improved transmission rate compared to existing coded caching schemes from t-designs for the same set of parameters $K, F$ and $Z$. Scheme II is also shown to be optimal in some cases. Both schemes require less subpacketization but large cache sizes. We then constructed two classes of HPDA from $t$-designs. The proposed HPDA constructions give low subpacketization level schemes with a larger mirror cache size and a low user cache size. 
\section*{Acknowledgement}
This work was supported partly by the Science and Engineering Research Board (SERB) of Department of Science and Technology (DST), Government of India, through J.C Bose National Fellowship to B. Sundar Rajan.

\section*{Appendix}
\subsection{Lower bounds on $S$ Comparison}
The lower bound on $S$ given in Theorem \ref{thm:MJXQ}
\begin{equation*}
	\begin{split}
		& = \left\lceil \frac{(F-Z)K}{F}\right\rceil + \left\lceil \frac{F-Z-1}{F-1} \left\lceil \frac{(F-Z)K}{F}\right\rceil \right\rceil \\ & + ...+ \left\lceil \frac{1}{Z+1} \left\lceil \frac{2}{Z+2} \left\lceil ...\left\lceil \frac{(F-Z)K}{F} \right\rceil ... \right\rceil \right\rceil \right\rceil \\ & \geq \left\lceil \frac{(F-Z)K}{F} + \frac{(F-Z-1)}{F-1} \frac{(F-Z)K}{F} \right.\\ & \left. + ...+ \frac{1}{Z+1} \frac{2}{Z+2} ...\frac{(F-Z)K}{F} \right\rceil \\ &
		= \left\lceil K\frac{(F-Z)!}{F!}\left\{ \frac{F!}{(F-Z)!}\frac{(F-Z)}{F} \right. \right.\\ & + \left. \left. \frac{F!}{(F-Z)!} \frac{(F-Z-1)}{F-1} \frac{(F-Z)}{F} + ...+ \right. \right. \\ & \left. \left. \frac{F!}{(F-Z)!}\frac{1}{F-(F-Z-1)} \frac{2}{F-(F-Z-2)} ...\frac{(F-Z)}{F} \right\} \right\rceil
	\end{split}
\end{equation*}
\begin{equation*}
	\begin{split} & 
		  = \left\lceil K\frac{(F-Z)!}{F!}\left\{ \frac{(F-1)!}{(F-Z-1)!} + \frac{(F-2)!}{(F-Z-2)!}+ ...+ \right. \right. \\ & \left. \left. \frac{(F-(F-Z))!}{1!} \right\} \right\rceil \\ &
		  = \left\lceil K\frac{(F-Z)!Z!}{F!}\left\{ \binom{F-1}{Z} + \binom{F-2}{Z}+ ...+ \right. \right. \\ & \left. \left. \binom{F-(F-Z)}{Z} \right\} \right\rceil	\\ &
		  =\left\lceil K\frac{(F-Z)!Z!}{F!}\binom{F}{Z+1} \right\rceil = \left\lceil K\frac{(F-Z)}{Z+1} \right\rceil \\ & = \text{the lower bound on S given in Theorem \ref{thm:Wei}}.
	\end{split} 
\end{equation*}	\hfill $\blacksquare$ 
\subsection{Proof of Theorem \ref{thm:scheme1} }
 For the sake of convenience, we continue with the representation of $(A\textbackslash\{B\cup x\})_\alpha$ for the non $\star$ entries.
 
By $(\ref{eq:cons1})$, for a given column $(A,B)$, a $\star$ appear in a row $x \in \mathcal{X}$ if and only if $x \notin A $ or $x \in B$. Therefore, the number of $\star s $ in a column, $Z= (v-k)+(k-i) = (v-i)$. Thus $C1$ of PDA definition holds.  $C2$ is obvious by construction.

Assume that $(A\textbackslash\{B\cup x\})_\alpha$ appears in the column $(A,B)$, then by $(\ref{eq:cons1})$, $x \in A \text{ and } x \notin B $. Therefore, $(B \cup x)$ will be different for each $x \in \mathcal{X}$. Thus, $(A\textbackslash\{B\cup x\})$ will be different in a given column  $(A,B)$. Thus   $(A\textbackslash\{B\cup x\})_\alpha$ appear only once in a given column. Since $\alpha$ denotes the $\alpha^{th}$ occurrence of $A\textbackslash\{B\cup x\}$ from left to right in the row indexed by $x$,  $(A\textbackslash\{B\cup x\})_\alpha$ cannot appear more than once in a row. Thus $C3.(a)$ of PDA definition holds.

For any two distinct entries $P_{x_i,(A_i,B_i)}$, $P_{x_j,(A_j,B_j)}$ with $x_i \neq x_j$ and $(A_i,B_i) \neq (A_j,B_j)$, let $ P_{x_i,(A_i,B_i)} = P_{x_j,(A_j,B_j)} = (A\textbackslash\{B\cup x\})_\alpha$. Then by $(\ref{eq:cons1})$, $A_i\textbackslash\{B_i\cup x_i\}=A_j\textbackslash\{B_j\cup x_j\}=A\textbackslash\{B\cup x\}$. Also $x_i \in A_i$, $x_i \notin B_i $, $x_j \in A_j$ and $x_j \notin B_j$. Therefore $x_i \notin A_i\textbackslash\{B_i\cup x_i\}$, which implies $x_i \notin A_j\textbackslash\{B_j\cup x_j\}$. That means, either  $x_i \notin A_j$ or $x_i \in \{B_j\cup x_j\}$. Thus  $x_i \notin A_j$ or $x_i \in B_j \text{ since } x_i \neq x_j$. Thus by  $(\ref{eq:cons1}), P_{x_i,(A_j,B_j)} = \star $. Similarly $P_{x_j,(A_i,B_i)} = \star$. Thus $C3.(b)$ of PDA definition holds. 
  
The regularity \textit{g} of the above PDA is obtained by finding in how many rows a given $(A\textbackslash\{B\cup x'\})$ appears. By $(\ref{eq:cons1})$,  $(A\textbackslash\{B\cup x'\})$ will not appear in those rows $x \in \mathcal{X}$ such that $x \in (A\textbackslash\{B\cup x'\})$. Also $|A\textbackslash\{B\cup x'\}|=(i-1)$, therefore  $g=v-(i-1)=(v-i+1)$.  Thus $C2'$ of g-PDA definition holds. Therefore, the array $\mathbf{P}$ is a $(v-i+1)-\left( \frac{\lambda\binom{v}{t}\binom{k}{i}}{\binom{k}{t}},v,v-i,\frac{\lambda\binom{v}{i-1}\binom{v-i}{t-i}}{\binom{k-i}{t-i}} \right)$ PDA. \hfill $\blacksquare$
	
\subsection{Proof of Theorem \ref{thm:scheme2} }
	For the sake of convenience, we continue with the representation of $(Y\textbackslash X)_\alpha$ for the non $\star$ entries.
	
By $(\ref{eq:cons2})$, for a given column $(A,Y)$, a $\star$ appears in a row $X \in \binom{\mathcal{X}}{i}$  if and only if $X \not\subset Y $. Number of $i$ sized subset ($X$) of $(i+1)$ sized set $Y$ is equal to $(i+1)$. Therefore, the number of $\star $s in a column, $Z= \binom{v}{i}-(i+1)$. Thus $C1$ of PDA definition holds.  $C2$ is obvious by construction. 

Assume that $(Y\textbackslash X)_\alpha$ appears in the  column $(A,Y)$, then by $(\ref{eq:cons2})$, $X \subset Y $ and $|Y\textbackslash X|=1$. Therefore, $(Y\textbackslash X)$ will be different for each $X$. Thus   $(Y\textbackslash X)_\alpha$ appears only once in a given column. Since $\alpha$ denotes the $\alpha^{th}$ occurrence of $Y\textbackslash X$ from left to right in the row indexed by $X$, $(Y\textbackslash X)_\alpha$ cannot appear more than once in a row. Thus $C3.(a)$ of PDA definition holds. 

For any two distinct entries $P_{X_i,(A_i,Y_i)}$, $P_{X_j,(A_j,Y_j)}$ with $X_i \neq X_j$ and $(A_i,Y_i) \neq (A_j,Y_j)$, let $ P_{X_i,(A_i,Y_i)} = P_{X_j,(A_j,Y_j)} = (Y\textbackslash X)_\alpha $. Then by $(\ref{eq:cons2})$, $Y_i\textbackslash X_i= Y_j\textbackslash X_j= Y\textbackslash X$. Also $ X_i\subset Y_i$  and $X_j \subset Y_j $. Therefore, $(Y_i \textbackslash X_i) \cup X_j = Y_j $. Since $X_i \neq X_j $, there exist at least one $x$ such that $x \in X_i \textbackslash X_j$, which implies $x \notin Y_i \textbackslash X_i$ and $x \notin X_j$. Therefore,  $x \notin \{(Y_i \textbackslash X_i) \cup X_j\}$ which implies $x \notin Y_j$. That means  $X_i \not\subset Y_j $. Therefore, by $(\ref{eq:cons2})$, $P_{X_i,(A_j,Y_j)}= \star $. Similarly, $P_{X_j,(A_i,Y_i)}= \star $. Thus $C3.(b)$ of PDA definition holds.
	
The regularity \textit{g} of the above PDA is obtained by finding in how many rows $(Y\textbackslash X)$ appears. By  $(\ref{eq:cons2})$, each $(Y\textbackslash X)$ will appear in those rows $X \in \binom{\mathcal{X}}{i}$ in which  $(Y\textbackslash X)$ does not belong. Also $|Y\textbackslash X|=1$. Therefore,  $g=\binom{v-1}{i}$.  Thus $C2'$ of g-PDA definition holds. 
	Therefore, the array $\mathbf{P}$ is a $
		\binom{v-1}{i}-\left( \frac{\lambda\binom{v}{t}\binom{k}{i+1}}{\binom{k}{t}},\binom{v}{i},\binom{v}{i}-(i+1),\frac{v\lambda\binom{v-(i+1)}{t-(i+1)}}{\binom{k-(i+1)}{t-(i+1)}} \right) \text{PDA}.
	$ \hfill $\blacksquare$ 	
\subsection{Proof of Theorem \ref{thm:opt_schm_1}}
	Scheme I gives a $(v-i+1)-\left( \frac{\lambda\binom{v}{t}\binom{k}{i}}{\binom{k}{t}},v,v-i,\frac{\lambda\binom{v}{i-1}\binom{v-i}{t-i}}{\binom{k-i}{t-i}} \right)$ PDA for any $i \in [1:t]$ from a $t-(v,k,\lambda)$ design. \\
For this PDA,
	\begin{equation*}
		\begin{split}
			& \left\lceil \frac{K(F-Z)}{Z+1} \right\rceil = \left\lceil  \frac{\frac{\lambda\binom{v}{t}\binom{k}{i}}{\binom{k}{t}}(v-(v-i))}{v-i+1} \right\rceil \\ & = \left\lceil \frac{\lambda v!(k-t)!i}{(v-t)!(k-i)!i!(v-i+1)} \right\rceil \\ & =\left\lceil \frac{v!i}{i!(v-i+1)(v-i)!} \frac{\lambda (k-t)!(v-i)!}{(v-t)!(k-i)!} \right\rceil \\ & = \left\lceil  \frac{v!}{(i-1)!(v-i+1)!}\frac{\lambda \binom{v-i}{t-i}}{\binom{k-i}{t-i}}  \right\rceil \\ & = \left\lceil \binom{v}{i-1}\frac{\lambda \binom{v-i}{t-i}}{\binom{k-i}{t-i}} \right\rceil = \left\lceil S \right\rceil = S.
		\end{split}
	\end{equation*}  
	 	That is, this class of PDA is an RPDA. Therefore this is an \textit{optimal PDA}. 
	 \hfill $\blacksquare$.
\subsection{Proof of Theorem \ref{thm:opt_schm_2}}
	Scheme II in \textit{Theorem 6} gives a  $\left( \frac{\lambda\binom{v}{t}\binom{k}{i+1}}{\binom{k}{t}},\binom{v}{i},\binom{v}{i}-(i+1),\frac{v\lambda\binom{v-(i+1)}{t-(i+1)}}{\binom{k-(i+1)}{t-(i+1)}} \right)$ PDA for any $i \in [1:t-1]$ from a $t-(v,k,\lambda)$ design. 
	
	For $i=1$, we obtain a $\left( \frac{\lambda\binom{v}{t}\binom{k}{2}}{\binom{k}{t}},v,v-2,\frac{v\lambda\binom{v-2}{t-2}}{\binom{k-2}{t-2}} \right)$ PDA. The parameters of this PDA are the same as that of Scheme I for $i=2$. Therefore, this PDA is an \textit{optimal PDA}.
	
	For $ \lambda =1$ and $i=t-1$, we obtain a $\left(\binom{v}{t},\binom{v}{t-1},\binom{v}{t-1}-t,v \right)$ PDA. For this PDA, $\frac{(F-Z)K}{F}=(v-t+1)$. Therefore, by $(\ref{eq:MJXQ})$, \\
	$S \geq (v-t+1) + \left\lceil \frac{t-1}{\binom{v}{t-1}-1} (v-t+1) \right\rceil  + ...+ \left\lceil \frac{1}{\binom{v}{t-1}-(t-1)} \left\lceil \frac{2}{\binom{v}{t-1}-(t-2)} \left\lceil ...(v-t+1) ... \right\rceil \right\rceil \right\rceil $. \\ Since there are $(t-1)$ terms after the first term $(v-t+1)$, $S \geq (v-t+1) + (t-1) = v $. That is, this PDA is a minimal load PDA. Therefore this is an \textit{optimal PDA}. 
		
	For $ \lambda=1$ and $i=t-1$,, from transpose of Scheme II, we obtain a $\left( \binom{v}{t-1},\binom{v}{t},\binom{v}{t}-(v-t+1),v \right)$ PDA. For this PDA, $\frac{(F-Z)K}{F}=\frac{(v-t+1)\binom{v}{t-1}}{\binom{v}{t}}=t$. Therefore, by $(\ref{eq:MJXQ})$, 
	$S \geq t + \left\lceil \frac{v-t}{\binom{v}{t}-1} t \right\rceil  + ...+ \left\lceil \frac{1}{\binom{v}{t}-(v-t)} \left\lceil \frac{2}{\binom{v}{t}-(v-t-1)} \left\lceil ...t ... \right\rceil \right\rceil \right\rceil $. \\
	Since there are $(v-t)$ terms after the first term $t$, \\ $S \geq (t + (v-t)) = v $. Therefore, the obtained PDA is an optimal PDA. \hfill $\blacksquare$
\subsection{Proof of Theorem \ref{thm:HPDA}}
	The proof follows from the HPDA construction described below. 
	\subsubsection{HPDA Construction} The HPDA in Theorem \ref{thm:HPDA} can be constructed from the PDA $\mathbf{P}$ constructed in Scheme I. Let each block $A \in \mathcal{A}$ in the $t$-design represent a mirror site. Then the PDA $\mathbf{P}$, constructed by replacing the non star entries of array in $(\ref{eq:cons1})$ by $(\alpha-1)\binom{v}{i-1}+f\left(A\textbackslash\{B\cup x\}\right)$,   can be partitioned into $K_1=|\mathcal{A}|=\frac{\lambda\binom{v}{t}}{\binom{k}{t}}$ parts by column. Let $\mathbf{P}=\left(\mathbf{P^{(1)}},\mathbf{P^{(2)}},....,\mathbf{P^{(K_1)}} \right)$. 
	\begin{itemize}
	\item Construction of $\mathbf{Q^{(0)}}$.
	We construct the $F \times K_1$ mirror placement array $\mathbf{Q^{(0)}}=(q_{x,k_1}^{(0)})_{x \in [\mathcal{X}], k_1 \in [K_1]}$ where
	\begin{equation}\label{eq:Mirror}
		q_{x,k_1}^{(0)}= \begin{cases}  \star & \text { if } P_{x,(A,B)}=\star \hspace{0.2cm} \forall B \subset A \\ null & \text { otherwise }\end{cases}.
	\end{equation}
	i.e., $q_{x,k_1}^{(0)}$ is a $ \star $ if the row $x$ of $\mathbf{P^{(k_1)}}$ is a star row.
	\item Construction of $\mathbf{Q^{(k_1)}}$.
	Since each $A \in \mathcal{A}$ represents a mirror site, the columns corresponding to a given $A$ represent the number of users attached to it. Therefore, from $(\ref{eq:cons1})$, the number of users attached to each mirror site, $K_2=\binom{k}{k-i}$. Then $\mathbf{Q^{(k_1)}}$ is constructed by replacing the star entries in the star rows of $\mathbf{P^{(k_1)}}$ by distinct integers which has no intersection with $[S]$. 
	\item Construction of $\mathbf{Q}$.
	We get an $F \times (K_1+K_1K_2)$ array by arranging $ \mathbf{Q^{(0)}}$ and $\mathbf{Q^{(1)}},....,\mathbf{Q^{(K_1)}}$ horizontally. i.e., $\mathbf{Q}= \left( \mathbf{Q^{(0)}},\mathbf{Q^{(1)}},....,\mathbf{Q^{(K_1)}} \right)$.
\end{itemize}

	Now we have to verify that the array $\mathbf{Q}$ satisfies the definition of HPDA.
	\subsubsection{HPDA Properties Verification}
	By $(\ref{eq:Mirror})$, $q_{x,k_1}^{(0)}$ is a $ \star $ if the row $x$ of $\mathbf{P^{(k_1)}}$ is a star row. The row $x$ of $\mathbf{P^{(k_1)}}$ is a star row if and only if $x \notin A$. Since each $A \in \mathcal{A}$ is of size $k$ and $x$ can take $v$ values, the number of star rows in each  $\mathbf{P^{(k_1)}}$ is $(v-k)$. Therefore, each column of $\mathbf{Q^{(0)}}$ has $Z_1=(v-k)$ stars. Thus $B1$ of HPDA definition holds.
	
	Since $\mathbf{P}$ is a PDA, each $\mathbf{P^{(k_1)}}$ is a PDA with the number of columns equal to the number of $(k-i)$ sized subsets of a $k$ sized block $A$. Therefore, $K_2=\binom{k}{k-i}$. By $(\ref{eq:cons1})$, the number of stars in a given column, excluding the stars in a star row of $\mathbf{P^{(k_1)}}$ is, $Z_2 = \left\{|x| : x \in A \text{ and } x \cap B \neq \emptyset \right\}=\left\{|x| : x \in  B \right\}=(k-i)$.  Each $\mathbf{Q^{(k_1)}}$ is constructed by replacing the star rows of $\mathbf{P^{(k_1)}}$ by distinct integers which has no intersection with $[S]$. Let $S_{k1}$ denote the integer set of $\mathbf{Q^{(k_1)}}$. Therefore, $\mathbf{Q^{(k_1)}}$ is a $\left(\binom{k}{k-i},v,(k-i),|S_{k1}|\right)$ PDA. Thus, $B2$ of HPDA definition holds.
	
	$S_m$ denotes the set of distinct integers replacing the star entries in the star rows of $\mathbf{P^{(k_1)}}$ and $q_{x,k_1}^{(0)}$ is a $ \star $ if the row $x$ of $\mathbf{P^{(k_1)}}$ is a star row.  Thus, $B3$ of HPDA definition holds. 
	
	Let $q_{j,k_2}^{(k_1)}=q_{j',k'_2}^{(k'_1)}=s$, where $k_1 \neq k'_1$. The $s \notin S_m$ since each integer of $S_m$ occurs exactly once. If $q_{j',k_2}^{(k_1)}=s'$ is an integer, then $s' \in S_m$ since  $\mathbf{P}$ is a PDA. If $q_{j',k_2}^{(k_1)}=s' \in S_m$, then by our construction 	$q_{j',k_1}^{(0)}=\star$. Thus, $B4$ of HPDA definition holds. 
	
From the construction in $(\ref{eq:cons1})$, it is clear that any subarray $\mathbf{P^{(k_1)}}$ contains atmost $\binom{k}{i-1}\frac{\lambda\binom{v-i}{t-i}}{\binom{k-i}{t-i}}$ distinct integers. Let $S'_{k1}$ denote this set of integers in $\mathbf{Q^{(k_1)}}$. Therefore, $S'_{k1}\leq \binom{k}{i-1}\frac{\lambda\binom{v-i}{t-i}}{\binom{k-i}{t-i}}$.  Let $S^{''}_{k1}$ denote the set of distinct integers in $\mathbf{Q^{(k_1)}}$ replacing the star rows of $\mathbf{P^{(k_1)}}$. There are  $Z_1$ star rows in $\mathbf{P^{(k_1)}}$ and each star row contain $K_2$ stars. Therefore, $ S^{''}_{k1}=\left[S+1+(k_1-1)Z_1K_2: S + k_1Z_1K_2\right]=\left[S+1+(k_1-1)(v-k)\binom{k}{k-i}: S + k_1(v-k)\binom{k}{k-i}\right]$. Thus the integer set of $\mathbf{Q^{(k_1)}}$ is given by, $
	S_{k1} = S'_{k1} + S^{''}_{k1} \leq \binom{k}{i-1}\frac{\lambda\binom{v-i}{t-i}}{\binom{k-i}{t-i}} + \left[S+1+(k_1-1)(v-k)\binom{k}{k-i}: S + k_1(v-k)\binom{k}{k-i}\right]$. 
And, $S_m=\left[ S+1 : S+K_1Z_1K_2\right] =\left[S+1 : S+\frac{\lambda\binom{v}{t}\binom{k}{k-i}(v-k)}{\binom{k}{t}}\right]$.	The set of integers $\underset{k_1=1}{\bigcup^{K_1} S_{k_1}}\textbackslash S_m$ is same as the set of integers $[S]$ in original PDA $\mathbf{P}$. Therefore by Theorem \ref{thm:hpda_thm}, the transmission loads \\ $R_1=\frac{\left|\underset{k_1=1}{\bigcup^{K_1} S_{k_1}}\right|-\left|S_m \right|}{F} = \frac{S}{F}= \frac{\lambda\binom{v}{i-1}\binom{v-i}{t-i}}{\binom{k-i}{t-i}v}$ and \\ $R_2=\max_{k_1 \in [K_1]}\left\{ \frac{|S_{k_1}|}{F}\right\} \leq \frac{\binom{k}{i-1}\frac{\lambda\binom{v-i}{t-i}}{\binom{k-i}{t-i}}+(v-k)\binom{k}{k-i}}{v}$. \\ Thus we obtain the HPDA mentioned in Theorem \ref{thm:HPDA}. \hfill $\blacksquare$

\end{document}